\documentclass{aastex}

\usepackage{epsfig}

\usepackage[caption = false]{subfig}

\usepackage{epsfig}
\usepackage{amsmath}
\usepackage{graphicx}
\usepackage{float}
\usepackage{amsfonts}
\usepackage{amssymb}
\usepackage{makeidx}
\usepackage{mathrsfs}

\def\lapprox{\lower.4ex\hbox{$\;\buildrel <\over{\scriptstyle\sim}\;$}}

\def\green{f_{_{\rm G}}}
\def\Ngreen{N_{_{\rm G}}}
\def\Atilde{\tilde A}
\def\Btilde{\tilde B}
\def\Ctilde{\tilde C}

\def\Ftilde{\tilde F}

\newcommand{\Ncool}{{{\mathscr N}}_{_{\rm G}}}

\newcommand{\begeq}{\begin{equation}}
\newcommand{\fineq}{\end{equation}}
\newcommand{\begeqarray}{\begin{eqnarray}}
\newcommand{\fineqarray}{\end{eqnarray}}

\newcommand{\gapprox}{\lower.4ex\hbox{$\;\buildrel
>\over{\scriptstyle\sim}\;$}}

\slugcomment{accepted for publication in ApJ}

\shorttitle{Particle Acceleration in Crab Pulsar}

\shortauthors{Kroon, Becker, Finke, \& Dermer}

\begin{document}

\title{ELECTRON ACCELERATION IN PULSAR-WIND TERMINATION SHOCKS:
AN APPLICATION TO THE CRAB NEBULA GAMMA-RAY FLARES}

\author{John J. Kroon}

\author{Peter A. Becker}

\affil{Department of Physics and Astronomy, George Mason University, Fairfax, VA 22030-4444, USA; jkroon@gmu.edu, pbecker@gmu.edu}

\author{Justin D. Finke}

\affil{Space Science Division, Naval Research Laboratory, Washington, DC
20375, USA; justin.finke@nrl.navy.mil}

\author{Charles D. Dermer}

\affil{Department of Physics and Astronomy, George Mason University, Fairfax, VA 22030-4444, USA; charlesdermer@outlook.com}

\begin{abstract}
The $\gamma$-ray flares from the Crab nebula observed by {\it AGILE} and {\it Fermi}-LAT reaching GeV energies and lasting several days challenge the standard models for particle acceleration in pulsar wind nebulae, because the radiating electrons have energies exceeding the classical radiation-reaction limit for synchrotron. Previous modeling has suggested that the synchrotron limit can be exceeded if the electrons experience electrostatic acceleration, but the resulting spectra do not agree very well with the data. As a result, there are still some important unanswered questions about the detailed particle acceleration and emission processes occurring during the flares. We revisit the problem using a new analytical approach based on an electron transport equation that includes terms describing electrostatic acceleration, stochastic wave-particle acceleration, shock acceleration, synchrotron losses, and particle escape. An exact solution is obtained for the electron distribution, which is used to compute the associated $\gamma$-ray synchrotron spectrum. We find that in our model the $\gamma$-ray flares are mainly powered by electrostatic acceleration, but the contributions from stochastic and shock acceleration play an important role in producing the observed spectral shapes. Our model can reproduce the spectra of all the {\it Fermi}-LAT and {\it AGILE} flares from the Crab nebula, using magnetic field strengths in agreement with the multi-wavelength observational constraints. We also compute the spectrum and duration of the synchrotron afterglow created by the accelerated electrons, after they escape into the region on the downstream side of the pulsar wind termination shock. The afterglow is expected to fade over a maximum period of about three weeks after the $\gamma$-ray flare.

\end{abstract}



\section{INTRODUCTION}

The Crab nebula has been a consistent emitter of electromagnetic radiation spanning the range from radio to $\gamma$-rays for decades. The radio emission may represent the continual cooling of relic electrons and positrons associated with the initial explosion (Volpi et al. 2008). However, the higher frequency emission, up to photon energies of $\sim 1$\,MeV, seems to be powered by the conversion of the kinetic energy in the ``cold'' pulsar wind, with bulk Lorentz factor $\sim 10^6$, into a ``hot'' distribution of accelerated electrons, at the pulsar-wind termination shock, which is a standing shock generated where the wind encounters the comparatively dense material in the outer region of the synchrotron nebula (Montani \& Bernardini 2014). We will refer to the electrons and positrons collectively as ``electrons'' throughout the remainder of the paper. The pulsar wind termination shock is located at radius $r_t \sim 10^{17}\,$cm, which is about an order of magnitude less than the size of the nebula itself (Rees \& Gunn 1974).

\subsection{Quiescent Emission}
\label{sec:QE}

Particle acceleration must be occurring at the termination shock in order to explain the entire energy range of the observed quiescent emission (Gaensler \& Slane 2006). Electrons accelerated at the shock diffuse and advect outward, into the synchrotron-emitting region of the nebula, located downstream from the termination shock (Kennel \& Coroniti 1984; Hester 2008). The dependence of the synchrotron radiation lifetime on the electron energy causes the high-energy component of the quiescent emission to be produced near the shock. Conversely, the radio emission is the most extended component, filling the synchrotron nebula out to a radius of $\sim 10^{18}\,$cm. The spin-down luminosity of the pulsar, $\sim 5 \times 10^{38}\,\rm erg\,sec^{-1}$, is converted into radiation with a remarkable efficiency, approaching $\sim 30$\% (Abdo et al. 2011).

Despite the success of the qualitative picture described above, the physical details of the electron acceleration process required to explain the observed quiescent emission are not very well understood. In particular, the standard model for first-order Fermi acceleration (also called diffusive shock acceleration) at the termination shock does not seem to be able to explain the shape of the electron energy distribution implied by the observed synchrotron spectra, due to two problems (Olmi et al. 2015; Komissarov 2013). The first is that the termination shock is necessarily relativistic, since the upstream Lorentz factor of the ``cold'' pulsar wind is $\Gamma_u \gg 1$ (Lyubarsky 2003; Aharonian et al. 2004), and the efficiency of relativistic shock acceleration is limited (Ellison et al. 1990). The second problem is that the magnetic field orientation at the termination shock is likely to be parallel to the shock front, and this reduces the probability of electrons to recycle back to the upstream side of the shock to experience further acceleration (Gallant et al. 1992; Sironi et al. 2015). Similar problems limit the efficiency of the shock-drift acceleration mechanism (Aharonian et al. 2004). These issues have led to the exploration of models for the quiescent emission that invoke particle acceleration driven by magnetic reconnection in the vicinity of the shock, but it is not yet clear whether this process provides a complete explanation (Coroniti 1990; Olmi et al. 2015).

Another puzzle related to the observed quiescent emission is the so-called ``$\sigma_{\rm mag}$ problem,'' where the magnetization parameter, $\sigma_{\rm mag}$, is defined by (e.g., Sironi \& Spitkovsky 2014)
\begin{equation}
\sigma_{\rm mag} \equiv \left(\frac{v_{\rm A}}{c}\right)^2 \ , \qquad
v_{\rm A} \equiv \frac{B}{\sqrt{4 \pi \rho}} \ ,
\label{eqAlfven}
\end{equation}
and $v_{\rm A}$ and $\rho$ denote the Alfv\'en velocity and the mass density in the upstream (cold) pulsar wind, respectively. We note that $\sigma_{\rm mag}$ can exceed unity, since the Alfv\'en velocity defined above can exceed $c$. However, in such cases, the phase velocity of the magnetohydrodynamical (MHD) waves is limited to $c$ (Heyvaerts et al. 2012). The problem is that the pulsar wind itself must be formed in a region with $\sigma_{\rm mag} \gg 1$ (Gaensler \& Slane 2006), whereas the acceleration of the electrons at the termination shock requires $\sigma_{\rm mag} \ll 1$ in order to satisfy the downstream boundary conditions in the synchrotron nebula (Rees \& Gunn 1974; Kennel \& Coroniti 1984).

\subsection{Gamma-Ray Flares}
\label{sec:GRFlares}

The formulation of a theoretical explanation for the quiescent emission from the Crab nebula already presents significant challenges, as discussed above, but a whole new set of problems emerged with the detection of a series of six remarkable short-duration $\gamma$-ray flares from 2010 September through 2013 March, five of which produced high-quality {\it Fermi}-LAT spectra. During these events, the Crab nebula brightened by about an order of magnitude in the $\sim 100-500\,$GeV energy range. At the peak of the ``super flare'' of 2011 April, the Crab nebula was the brightest $\gamma$-ray source in the sky, with an observed flux exceeding that of the Vela pulsar, or any of the $\gamma$-ray emitting active galaxies. The spectral shape of the 2011 April transient was a broad hump in the energy range 0.1-1\,GeV, with a rapid increase over a few days, from April 13-15, and a total duration of about 9 days (Buehler at al. 2012).

At its brightest, the $\gamma$-ray luminosity of the 2011 April flare corresponds to $\sim 1\,$\% of the spin-down power of the rotating neutron star, if emitted isotropically. This is about 30 times the average value for the Crab nebula. The second-brightest flare was observed in 2007 September (Kargaltsev et al. 2015; Buehler \& Blandford 2014), and the statistics were similar to the 2011 flare, including evidence for sub-day variability (see Striani et al. 2013, and references therein). These observation of $\gamma$-rays from the Crab nebula with energies at least an order of magnitude above the radiation-reaction limit, $\sim 200\,$MeV, presents serious challenges to the standard astrophysical particle acceleration mechanisms.

The observed transient GeV emission from the Crab nebula implies the presence of intense particle acceleration in the vicinity of the pulsar wind termination shock. The very large value of the upstream Lorentz factor $\Gamma_u \gg 1$ implies that relativistic shock acceleration may play an important role in the formation of the high energy $\gamma$-rays observed by {\it Fermi}-LAT and {\it AGILE} (see Buehler \& Blandford 2014, and references therein). Diffusive acceleration of electrons at a standing shock, whether relativistic or non-relativistic, is mediated by MHD waves, and therefore the maximum acceleration rate is limited to the Bohm rate (Lemoine \& Waxman 2009). However, relativistic shocks can be less efficient accelerators than non-relativistic shocks, once the increase in the scattering time in the relativistic case is included (Sironi et al. 2015). Hence, one finds that the particle acceleration rate is limited to the Bohm rate whether the acceleration occurs via the first-order Fermi process operating at a shock, or via the second-order Fermi process due to stochastic wave-particles interactions. In either case, the maximum particle energy that can be achieved is ultimately limited by synchrotron losses, and the value of the maximum energy is obtained by equating the Bohm acceleration rate with the synchrotron loss rate. This yields the radiation-reaction, or ``synchrotron burnoff,'' limit for the electron energy (see Section~\ref{sec:synchb}). In the end, the limited efficiency of shock acceleration and stochastic wave-particle acceleration leads to the conclusion that these processes are unable to explain the observed high-energy transient emission from the Crab nebula, even when one includes the mild Doppler boost that occurs on the downstream side of the shock, unless the inclination angle of the pulsar is $\sim 90^\circ$ (Komissarov 2013).

A number of authors have attempted to explain the observed GeV transients from the Crab nebula by circumventing the synchrotron burnoff limit using a variety of physical mechanisms. This limit can be violated (1) for emission regions with bulk relativistic motion (e.g., Lyutikov et al. 2012), (2) by acceleration in a low magnetic-field region and radiation in a high-field region (e.g., Komissarov \& Lyutikov 2011), and (3) if an accelerating electric field is present, as produced, for example, by magnetic reconnection (e.g., Cerutti et al. 2012).

Significant progress has been made on scenario (3) using particle-in-cell (PIC) simulations (Cerutti et al. 2013, 2014a, 2014b), but a complete understanding of the particle acceleration phenomenon occurring in the Crab nebula pulsar wind is still lacking. In particular, it is not clear whether the level of magnetic suppression required in the reconnection models can be achieved. We will also focus on possibility (3) here, but we will use an alternative approach, based on mathematical analysis of a particle transport equation that includes terms describing electrostatic acceleration, stochastic and shock acceleration, shock-regulated (advective) escape, and escape from the nebula via Bohm diffusion. An exact solution is obtained for the resulting electron distribution. In the next section, we review some of the key observational diagnostics, and discuss the associated requirements and constraints for the theoretical models.

\subsection{Flare Energetics}
\label{sec:FE}

The characteristic peak synchrotron energy emitted by an isotropic distribution of relativistic electrons with Lorentz factor $\gamma$ spiraling in a magnetic field with strength $B$ is (e.g., Rybicki \& Lightman 1979)
\begin{equation}
\epsilon_{\rm pk}(\gamma) = \frac{B}{B_{\rm crit}} \, \gamma^2 m_e c^2
= 231.5 \ {\rm MeV} \, \left(\frac{\gamma}{10^{10}}\right)^2 \left(\frac{B}{200 \, \mu {\rm G}}\right)
\label{eq1newnew}
\ ,
\end{equation}
where $m_e$ is the electron mass, $c$ is the speed of light, and the critical magnetic field, $B_{\rm crit}$, is defined by
\begin{equation}
B_{\rm crit} \equiv \frac{2\pi m_e^2 c^3}{e h} = 4.41 \times 10^{13}
\ \rm G \ .
\label{eq2newnew}
\end{equation}
Observational estimates of the magnetic field strength in the Crab nebula are typically close to $B \sim 200 \ \mu$G (e.g., Aharonian et al. 2004). The generation of synchrotron emission with an energy of $\sim 1\,$GeV by an isotropic electron distribution in the presence of such a field therefore requires a Lorentz factor $\gamma \sim
2 \times 10^{10}$.

The synchrotron energy loss rate per electron, averaged over an isotropic distribution of pitch angles, is given by
\begin{equation}
<\!\dot\gamma\!>_{\rm syn} m_e c^2 = - \frac{4}{3}\gamma^2
\sigma_{\rm T}c\frac{B^2}{8\pi} \ ,
\label{eq4newnew}
\end{equation}
where $\sigma_{\rm T}$ denotes the Thomson cross section. The associated synchrotron momentum loss rate can be written as
\begin{equation}
<\!\dot p\!>_{\rm syn} = - \frac{B_0}{m_e c} p^2 \ ,
\label{eq18newnew}
\end{equation}
where the constant $B_0 \propto \rm s^{-1}$ is given by
\begin{equation}
B_0 = \frac{\sigma_{\rm T} B^2}{6 \pi m_e c} \ .
\label{eq19newnew}
\end{equation}

The synchrotron lifetime of the relativistic electrons producing the flare can be estimated using
\begin{equation}
t_{\rm syn} = - \frac{\gamma}{<\!\dot\gamma\!>_{\rm syn}} \ ,
\label{eq3newnew}
\end{equation}
which can be combined with Equation~(\ref{eq4newnew}) to obtain
\begin{equation}
t_{\rm syn} = \frac{6 \pi m_e c^2}{\sigma_{\rm T} c B^2 \gamma}
= 22.4 \ {\rm days} \ \left(\frac{B}{200 \, \mu {\rm G}}\right)^{-2}
\left(\frac{\gamma}{10^{10}}\right)^{-1}
\ .
\label{eq5newnew}
\end{equation}
We can also express the synchrotron lifetime as a function of the peak flare photon energy, $\epsilon_{\rm pk}$, by using Equation~(\ref{eq1newnew}) to eliminate the Lorentz factor $\gamma$ in Equation~(\ref{eq5newnew}),
obtaining
\begin{equation}
t_{\rm syn} = 10.8 \ {\rm days} \ \left(\frac{B}{200 \, \mu {\rm G}}
\right)^{-3/2} \left(\frac{\epsilon_{\rm pk}}{\rm GeV}\right)^{-1/2}
\ .
\label{eq6newnew}
\end{equation}
For a peak photon energy $\epsilon_{\rm pk} \sim 1\,$GeV, and field strength $B \sim 200\,\mu$G, we obtain roughly the observed flare duration, which supports the interpretation that the observed $\gamma$-ray flares are the result of synchrotron emission (e.g., Abdo et al. 2011). The similarity between the two timescales also suggests that the particle distribution during the peak of the flare is close to equilibrium. We will use this to motivate our consideration of the steady-state transport equation introduced in Section~\ref{sec:sste}. However, the issue of time dependence will be revisited in Section~9.3.

\subsection{Synchrotron Burnoff}
\label{sec:synchb}

The classical radiation-reaction (synchrotron burnoff) limit places severe constraints on the particle acceleration mechanism required to power the observed emission, as discussed below. Variability on shorter timescales ($\sim 1\,$day) may have also been observed (Buehler et al. 2012; Mayer et al. 2013; Striani et al. 2013), perhaps as a consequence of instabilities in the structure of the termination shock and the strength of the associated magnetic field.

The minimum timescale for the acceleration of relativistic electrons via energetic collisions with MHD waves is the Larmor timescale,
\begin{equation}
t_{\rm L} = \frac{r_{\rm L}}{c} = \frac{m_e c}{q B} \, \gamma
\ ,
\label{eq7newnew}
\end{equation}
where $r_{\rm L}$ is the Larmor radius and $q$ denotes the magnitude of the electron charge. The synchrotron burnoff limit is obtained by equating the Larmor timescale with the synchrotron loss timescale, given by Equation~(\ref{eq5newnew}), which yields an expression for the maximum Lorentz factor, $\gamma_{_{\rm MHD}}$, that can be achieved via MHD wave acceleration in the presence of a magnetic field of strength $B$, neglecting other factors. The result obtained is
\begin{equation}
\gamma_{_{\rm MHD}} = \sqrt{\frac{6\pi q} {B \sigma_{\rm T}}}
= 8.25 \times 10^9 \, \left(\frac{B}{200\,\mu\rm G}\right)^{-1/2}
\label{eq8newnew}
\ .
\end{equation}
We can substitute Equation~(\ref{eq8newnew}) into Equation~(\ref{eq1newnew}) to obtain the radiation-reaction-limited peak synchrotron energy, given by
\begin{equation}
\epsilon_{_{\rm MHD}} \equiv
\epsilon_{\rm pk}(\gamma_{_{\rm MHD}}) = \frac{6 \pi q m_e c^2}
{B_{\rm crit} \, \sigma_{\rm T}} = 158 \ {\rm MeV}
\ .
\label{eq9newnew}
\end{equation}
This is far below the highest energy observations which are in excess of 1\,GeV. Hence, the synchrotron burnoff limit implies that particle acceleration via interaction with MHD waves is insufficient to explain the energetics of the observed Crab nebula $\gamma$-ray flares.

The observation of $\gamma$-rays with energies exceeding the radiation-reaction limit given by Equation~(\ref{eq9newnew}) has motivated speculation that the radiating electrons are accelerated electrostatically in a magnetic reconnection region where electric fields are induced (e.g., Buehler at al. 2012; Cerutti et al. 2012). However, numerical simulations based on electrostatic acceleration in a magnetic reconnection region require the presence of a $\sim 5\,$mG magnetic field and PeV electrons (Cerutti et al. 2012, 2014a). This field strength is substantially higher than that indicated by the multi-wavelength observations of the quiescent spectrum, which suggest that the ambient field has strength $B \sim 200\,\mu$G (e.g., Aharonian et al. 2004; Meyer \& Horns 2010). These studies focused on the comprehensive broad-band (from radio through hard $\gamma$-rays) spectrum to infer the average ambient magnetic field responsible for the observed synchrotron emission from the Crab nebula.

In this paper, we seek to develop a self-consistent theoretical framework that can account for all of the Crab nebula $\gamma$-ray flare spectra detected by {\it Fermi}-LAT, based on the standard ambient magnetic field strength $B \sim 200\,\mu$G. Our model is based on a one-zone electron transport equation that includes terms describing particle injection, stochastic acceleration, electrostatic acceleration, shock acceleration, radiative losses, and particle escape. The model should be interpreted as a spatial average over the acceleration and emission regions, which may either be co-located or separate regions. The transport equation is solved to obtain a closed-form expression for the energy distribution of the relativistic electrons, which is then used to compute the $\gamma$-ray spectrum produced via direct synchrotron emission.

The remainder of the paper is organized as follows. In Section~\ref{sec:ptf}, we provide an overview of the physical background and develop the specific terms to be employed in the fundamental electron transport equation. In Section~\ref{sec:sste} we introduce and discuss the associated steady-state Fokker-Planck equation. In Section~\ref{sec:pdars} we obtain the exact solution for the electron Green's function, and we develop expressions for computing the corresponding synchrotron spectra. In Section~\ref{sec:crabapp} we apply the model to the transient $\gamma$-ray emission observed from the Crab nebula, and in Section~\ref{sec:afterglow} we compute and discuss the afterglow spectra. The flare energy budgets are analyzed and discussed in Section~\ref{sec:EnergyBudget}. In Section~\ref{sec:parcon} we examine the parameter constraints for the model, and in Section~\ref{sec:discuss} we review our main conclusions and discuss their astrophysical significance.

\section{PARTICLE TRANSPORT FORMALISM}
\label{sec:ptf}

The $\gamma$-ray flares observed from the Crab nebula significantly exceed the classical radiation-reaction limit, and therefore a natural explanation could be that the radiating electrons are accelerated, at least in part, by an electric field created in a region of magnetic reconnection on the downstream side of the termination shock (e.g., Komissarov 2013). Although a variety of numerical simulations have been employed to model this phenomenon, they have not been entirely successful at explaining the shape of the observed $\gamma$-ray spectra, and furthermore, they tend to invoke magnetic field strengths that are somewhat higher than those implied by observations of the Crab nebula.

The uncertainties regarding the numerical simulations have motivated us to revisit the problem using an analytical approach based on a transport equation that includes terms describing electrostatic acceleration, stochastic acceleration, shock acceleration, synchrotron losses, and particle escape. Since synchrotron losses are included in the transport equation, the subsequent calculation of the $\gamma$-ray synchrotron spectrum is self-consistent. By describing the particle distribution, and the resulting radiation spectrum, using analytical expressions, we are able to study of a wide range of parameter values, while maintaining complete control over the relative importance of the various physical processes. Conversely, complex numerical simulations can sometimes make it difficult to pinpoint the specific effect of each process, and they are usually not amenable to fitting observational data. Another advantage of closed-form analytical models it that they can often be evaluated in real-time, allowing one to perform quantitative fits to the observational data, and they are also useful for benchmarking more sophisticated numerical simulations. The similarity of the flare duration and the synchrotron lifetime given by Equation~(\ref{eq6newnew}) suggests that the electron distribution achieves an approximate equilibrium during the peak of the flare. We will use this fact to justify our utilization of a steady-state transport equation when we solve for the electron distribution in Section~\ref{sec:pdars}.

The relativistic electrons ejected from the central pulsar travel outward in the form of a cold, relativistic wind, with bulk Lorentz factor $\sim 10^6$. Since the particles propagate along the field lines, with a negligible transverse component, very little synchrotron emission is generated in the cold wind, and the particles are mainly subject to adiabatic losses (Kennel \& Coroniti 1984). Most of the observed radio emission is produced beyond the pulsar wind termination shock, which is located at radius $r_t \sim 10^{17}\,$cm. The spatial distribution of the radiation implies the existence of an efficient acceleration mechanism operating at the termination shock. An extension of this model into the $\gamma$-ray region has naturally led to the suggestion that the $\gamma$-rays emitted during the Crab nebula flares may also be produced in the vicinity of the shock (e.g. Begelman 1998; Komissarov \& Lyubarsky 2004; Uzdensky et al. 2011). The particle acceleration required to power the $\gamma$-ray flares may occur via a variety of mechanisms, including first-order Fermi acceleration due to multiple shock crossings, second-order Fermi acceleration due to stochastic interactions with a random field of MHD waves, and direct electrostatic acceleration in the electric field generated in the magnetic reconnection region surrounding the shock. In this section, we derive the spatial and momentum diffusion coefficients describing the stochastic interaction with the random field of MHD waves.

The strongest possible MHD acceleration occurs in the Bohm regime, when the particle mean-free path, $\ell$, is comparable to the Larmor radius (Krall \& Trivelpiece 1986),
\begin{equation}
r_{\rm L} = \frac{p c}{q B} \ ,
\label{eq14newnew}
\end{equation}
where $c$ is the speed of light, $q$ is the magnitude of the electron charge, and $B$ is the local magnetic field strength. It is convenient to parameterize the mean-free path relative to the Larmor radius by introducing the dimensionless parameter $\eta$, where
\begin{equation}
\eta \equiv \frac{\ell}{r_{\rm L}} \ .
\label{eqMFP}
\end{equation}
The value of $\eta$ determines the diffusion regime in which we are operating. In an ordered magnetic field, the Larmor radius represents the minimum mean-free path for interactions with MHD waves, so that $\eta \gapprox 1$, with the limiting case corresponding to Bohm diffusion. We develop an approximate quantitative expression for $\eta$ below.

Focusing on the Bohm diffusion limit, in an ordered field, the differential probability of scattering with an MHD wave in the distance interval $dx$ is given by $dP_{\rm L} = dx/r_{\rm L}$. However, in a disordered field, the particles also experience large-angle deflections due to encounters with stationary kinks, with a mean spacing equal to the magnetic coherence length, $\ell_{\rm coh}$. The differential probability of deflection due to an encounter with a kink in the distance interval $dx$ is $dP_{\rm coh} = dx/\ell_{\rm coh}$. Taking both Bohm diffusion and magnetic deflection into consideration, the total scattering probability per unit length, $dP_{\rm tot}$, is given by the sum
\begin{equation}
dP_{\rm tot} = dP_{\rm L} + dP_{\rm coh} \ .
\label{eqScattProb}
\end{equation}
Substituting for $dP_{\rm L}$ and $dP_{\rm coh}$, and setting $dP_{\rm tot} = dx/\ell$, we find that the effective mean-free path $\ell$ is related to $r_{\rm L}$ and $\ell_{\rm coh}$ via
\begin{equation}
\frac{1}{\ell} = \frac{1}{r_{\rm L}} + \frac{1}{\ell_{\rm coh}} \ ,
\label{eqMeanEll}
\end{equation}
and the corresponding expression for $\eta$ is therefore (see Equation~(\ref{eqMFP}))
\begin{equation}
\eta = \left({1+ \frac{r_{\rm L}}{\ell_{\rm coh}}}\right)^{-1} \ .
\label{eqEtaExp}
\end{equation}
This result is reasonably consistent with the mean-free path variation obtained by Zank et al. (2004) using numerical simulations of charged particle diffusion in magnetized plasmas.

In general, one expects that $\ell/r_{\rm L} \sim B/\delta B$, where $\delta B$ is the turbulent component of the magnetic field (Hussein \& Shalchi 2014; Giacalone \& Jokipii 1999; Dosch et al. 2011). Hence, in a turbulent field with $\delta B \sim B$, it follows that $\eta \sim 1$, and that is the situation assumed here. This is consistent with Equation~(\ref{eqEtaExp}) provided the coherence length of the magnetic field is not much smaller than the Larmor radius for the particles emitting the $\gamma$-rays, which would imply that $\eta \ll 1$. Once the value of $\eta$ is specified, the associated spatial diffusion coefficient, $\kappa$, for diffusion with mean-free path $\ell = \eta r_{\rm L}$ is computed using (Dr\"oge \& Schlickeiser 1986; Reif 1965)
\begin{equation}
\kappa = \frac{c\, \ell}{3} = \eta \frac{c r_{\rm L}}{3} \ .
\label{eq15newnew}
\end{equation}

\subsection{Particle Distribution and Transport Equation}
\label{sec:PDandTE}

The spatial transport of the electrons in the environment surrounding the pulsar wind termination shock is governed by a combination of advection in the wind and spatial diffusion relative to the wind. In the Crab nebula, the particle acceleration and the production of the observed $\gamma$-rays may occur in separate geometrical regions (e.g., Cerutti et al. 2012), but it is not clear whether this has to be the case. Therefore, in the present paper, we will develop a one-zone spatial model, which represents an average over the acceleration and emission regions. In this approach, the spatial aspects of the problem are modeled using a simple escape-probability formalism, with escape timescale $t_{\rm esc}$. The energy dependence of $t_{\rm esc}$ depends on the nature of the mechanism transporting electrons out of the acceleration region. In our application, the dominant escape mechanism is expected to be a combination of shock-regulated (advective) escape on small scales, and Bohm diffusion on large scales, as discussed in Section~\ref{sec:pe}.

Based on these physical considerations, the transport equation we will utilize to model the evolution of the relativistic electron momentum distribution, $f$, in the pulsar wind nebula is given by (e.g., Becker et al. 2006; Park \& Petrosian 1995)
\begin{equation}
\frac{\partial f}{\partial t} = - \frac{1}{p^2}\frac{\partial}{\partial
p}\Big\{p^2\Big[-D(p)\frac{\partial f}{\partial p}
+ <\!\dot p\!>_{\rm gain}f + <\!\dot p\!>_{\rm loss}f\Big]\Big\}-
\frac{f}{t_{\rm esc}(p)} + \dot f_{\rm source}(p)
\ ,
\label{eq10newnew}
\end{equation}
where $p$ is the particle momentum, and the terms on the right-hand side represent stochastic (second-order) Fermi acceleration (i.e., momentum diffusion); systematic gains due to electrostatic acceleration and first-order Fermi acceleration at the shock; systematic losses due to synchrotron emission; particle escape; and particle injection, respectively. The distribution function $f$ is related to the total number of electrons, $N_e$, via
\begin{equation}
N_e(t) = \int_0^\infty 4\pi p^2 f(p,t) \, dp
\label{eq11newnew}
\ .
\end{equation}

\subsection{Stochastic Acceleration}
\label{sec:smhd}

The general relation between the spatial diffusion coefficient $\kappa$ and the momentum diffusion coefficient $D$ can be written as (Dr\"oge et al. 1987; Schlickeiser 1985)
\begin{equation}
D(p) \kappa(p) = \frac{p^2 v_{\rm A}^2}{9} \ ,
\label{eq16newnew}
\end{equation}
where $v_{\rm A}$ denotes the Alfv\'en velocity. Combining Equations~(\ref{eq14newnew}), (\ref{eq15newnew}), and (\ref{eq16newnew}), we find that in the Bohm scenario, the momentum dependence of $D(p)$ is given by
\begin{equation}
D(p) = D_0 m_e c \, p
\label{eq17newnew}
\ ,
\end{equation} 
where the constant $D_0$, with units of $\rm s^{-1}$, is defined by
\begin{equation}
D_0 \equiv \frac{q B \sigma_{\rm mag}}{3 \eta m_e c}
= 1172 \ {\rm s}^{-1} \, \sigma_{\rm mag}
\left(\frac{B}{200\,\mu{\rm G}}\right) \ \eta^{-1}
\ ,
\label{eq17Bnewnew}
\end{equation}
and the magnetization parameter, $\sigma_{\rm mag}$, is related to the Alfv\'en velocity via Equation~(\ref{eqAlfven}). Since $D(p) \propto p$ in Equation~(\ref{eq17Bnewnew}), this situation corresponds to the wave index $q=1$ case discussed by Becker et al. (2006) and Dermer et al. (1996).

As discussed in Section~\ref{sec:synchb}, the MHD acceleration timescale must exceed the gyroperiod of the accelerated electrons. We can use this fact to develop a quantitative restriction on the value of the momentum diffusion rate constant, $D_0$. We note that the mean stochastic momentum gain rate is given by (e.g., Becker et al. 2006)
\begin{equation}
<\!\dot p\!>_{\rm stoch} = \frac{1}{p^2} \frac{\partial}{\partial p}
\left[p^2 D(p)\right] = 3 D_0 m_e c \ ,
\label{eq75}
\end{equation}
where the final result follows from Equation~(\ref{eq17newnew}). The theoretical equilibrium momentum for a combination of synchrotron losses and stochastic acceleration, $p_{\rm stoch}=m_e c \, \gamma_{\rm stoch}$, neglecting all other processes, is computed by equating the stochastic gain rate, $<\!\dot p\!>_{\rm stoch}$, with the synchrotron loss rate, $<\!\dot p\!>_{\rm syn}$, so that
\begin{equation}
\Big(<\!\dot p\!>_{\rm stoch} + <\!\dot p\!>_{\rm syn}\Big)
\bigg|_{p=p_{\rm stoch}} = 0 \ .
\label{eqStochEquil1}
\end{equation}
We can combine this expression with Equations~(\ref{eq18newnew}) and (\ref{eq75}) to obtain
\begin{equation}
\Big(3 D_0 m_e c - \frac{\sigma_{\rm T} B^2 \gamma^2}{6 \pi}\Big)
\bigg|_{\gamma=\gamma_{\rm stoch}} = 0 \ ,
\label{eqStochEquil2}
\end{equation}
and therefore the theoretical equilibrium Lorentz factor for stochastic acceleration is given by
\begin{equation}
\gamma_{\rm stoch} = \frac{p_{\rm stoch}}{m_e c}
= \sqrt{\frac{18 \pi D_0 m_e c}{\sigma_{\rm T} B^2}}
\ ,
\label{eqStochEquil3}
\end{equation}
or, equivalently,
\begin{equation}
\gamma_{\rm stoch} = 2.41 \times 10^9 \left(\dfrac{B}{200 \, \mu {\rm G}}\right)^{-1}
\left(\dfrac{D_0}{100}\right)^{1/2}
\ .
\label{eqStochEquil3b}
\end{equation}

The synchrotron burnoff limit requires that $\gamma_{\rm stoch}$ cannot exceed $\gamma_{_{\rm MHD}}$ given by Equation~(\ref{eq8newnew}), and therefore we obtain the condition
\begin{equation}
D_0 \le D_0^{\rm max} \ ,
\label{eq79}
\end{equation}
where
\begin{equation}
D_0^{\rm max} \equiv
\frac{q B}{3 m_e c} = 1172 \ {\rm s}^{-1}
\left(\frac{B}{200\,\mu{\rm G}}\right)
\ .
\label{eq80}
\end{equation}
By combining Equations~(\ref{eq17Bnewnew}) and (\ref{eq80}), we find that in our model, the relationship between $D_0$ and $D_0^{\rm max}$ is given by
\begin{equation}
\frac{D_0}{D_0^{\rm max}} = \frac{\sigma_{\rm mag}}{\eta}
\ ,
\label{eq80b}
\end{equation}
which implies that the value of $D_0$ computed using Equation~(\ref{eq17Bnewnew}) will automatically satisfy the synchrotron burnoff constraint given by Equation~(\ref{eq79}), provided $\eta \sim 1$ and $\sigma_{\rm mag} \lesssim 1$, which are both reasonable expectations in the Crab pulsar wind nebula.

\subsection{Shock Acceleration}
\label{sec:ShockAcc}

In addition to the second-order Fermi acceleration discussed in Section~\ref{sec:smhd}, resulting from stochastic wave-particle interactions, the electrons also experience a combination of electrostatic and shock acceleration in the vicinity of the termination shock (Abdo et al. 2011). We discuss these two processes in this section.

The shape of the electron energy distribution resulting from first-order Fermi acceleration at the pulsar wind termination shock depends on the amount of time the particles spend in the acceleration region, which is regulated by the combined action of spatial diffusion and advection. On the upstream side of the termination shock, the wind is ultrarelativistic, with bulk Lorentz factor $\Gamma_u \sim 10^3-10^6$ (Kennel \& Coroniti 1984; Lyubarsky 2003; Aharonian et al. 2004). On the downstream side of the shock, the flow is mildly relativistic, with speed $c/3$ and bulk Lorentz factor $\Gamma_d \sim 1.1$ in the shock frame (Achterberg et al. 2001). In studies of particle acceleration in super-driven relativistic blast waves, which is the typical application of the theory, the Lorentz transformation between the shock frame and the frame at infinity creates a large deviation between the time and energy measured in the two frames. However, in our application, the pulsar wind termination shock is a stationary phenomenon, and therefore the shock frame is equivalent to the frame at infinity. Hence in the shock frame, the acceleration and cycle times can be estimated by writing (Lagage \& Cesarsky 1983)
\begin{equation}
t_{\rm accel} = t_{\rm cyc} = \frac{r_{\rm L}}{c} \ ,
\label{eq23newnew}
\end{equation}
where $r_{\rm L}$ is the Larmor radius (see Equation~(\ref{eq14newnew})).

Particles crossing the shock for the first time will experience an energy gain on the order of $\Gamma_u^2$ in the frame of the upstream gas, but subsequent crossings will increase the particle energy by a much smaller factor due to the dynamics of escape and acceleration (Achterberg et al. 2001). Ellison et al. (1990) find that the overall efficiency of particle acceleration at a relativistic shock is not much higher than that for a non-relativistic shock. The efficiency can be further reduced in the magnetic field is parallel to the shock front (Lemoine \& Waxman 2009), which may be the case in pulsar-wind termination shocks (Gallant et al. 1992; Sironi et al. 2015). Hence we will estimate the first-order Fermi acceleration rate experienced by the electrons due to multiple shock crossings in the laboratory (shock) reference frame by writing (Dermer \& Menon 2009)
\begin{equation}
<\!\dot p\!>_{\rm sh} = \xi \frac{p}{t_{\rm accel}}
= \xi q B
\ ,
\label{eqpdot}
\end{equation}
where the final (constant) result follows from Equations~(\ref{eq14newnew}) and (\ref{eq23newnew}), and $\xi$ is an efficiency factor reflecting the fact that the particle acceleration rate cannot exceed the Bohm rate, even in a relativistic shock (Lemoine \& Waxman 2009).

The theoretical equilibrium momentum for a combination of synchrotron losses and shock acceleration, $p_{\rm sh}=m_e c \, \gamma_{\rm sh}$, is obtained by balancing $<\!\dot p\!>_{\rm sh}$ with the synchrotron loss rate, $<\!\dot p\!>_{\rm syn}$, so that
\begin{equation}
\Big(<\!\dot p\!>_{\rm sh} + <\!\dot p\!>_{\rm syn}\Big)
\bigg|_{p=p_{\rm sh}} = 0 \ ,
\label{eqShockEquil1}
\end{equation}
which can be combined with Equations~(\ref{eq18newnew}) and (\ref{eqpdot}) to obtain
\begin{equation}
\Big(\xi q B - \frac{\sigma_{\rm T} B^2 \gamma^2}{6 \pi}\Big)
\bigg|_{\gamma=\gamma_{\rm sh}} = 0 \ .
\label{eqShockEquil2}
\end{equation}
The theoretical equilibrium Lorentz factor for shock acceleration versus synchrotron losses (neglecting all other processes) is therefore given by
\begin{equation}
\gamma_{\rm sh} = \frac{p_{\rm sh}}{m_e c}
= \sqrt{\frac{6 \pi \xi q}{\sigma_{\rm T} B}}
\ ,
\label{eqShockEquil3}
\end{equation}
or, equivalently,
\begin{equation}
\gamma_{\rm sh} = 8.25 \times 10^9 \left(\frac{B}{200 \, \mu {\rm G}}\right)^{-1/2} \xi^{1/2} \ .
\label{eqShockEquil3b}
\end{equation}
The maximum value for the efficiency factor $\xi$ is obtained by ensuring that $\gamma_{\rm sh}$ does not exceed the synchrotron burnoff limit, $\gamma_{_{\rm MHD}}$, given by Equation~(\ref{eq8newnew}). Hence we obtain the constraint
\begin{equation}
\xi \le 1 \ .
\label{eqbetaeff2}
\end{equation}

It is interesting to compare the result for $\gamma_{\rm sh}$ given by Equation~(\ref{eqShockEquil3b}) with the corresponding maximum Lorentz factor obtained by Kennel \& Coroniti (1984). In their model, the particle distribution is not computed from first principles, and instead, they assume a power-law shape for the downstream electron distribution, with the upper limit for the Lorentz factor determined based on the requirement of confinement within the termination shock radius. They find that $\gamma \lapprox 2.5 \times 10^9$ for $\sigma_{\rm mag} = 0.001$, which is in reasonable agreement with the result we obtain by setting $\xi=0.1$ in Equation (\ref{eqShockEquil3b}).

It is important to note that in shock acceleration models, such as the one considered here, the electrons can attain Lorentz factors much higher than the mean value expected based on the thermalization of the incoming particles at a classical collisional shock. In the classical case, the bulk kinetic energy in the cold upstream wind, with Lorentz factor $\Gamma_u \approx 10^6$, is randomized and shared amongst all of the particles. This results in the deceleration of the flow, as the upstream bulk kinetic energy is converted into thermal energy, with a mean downstream (random) Lorentz factor $\sim 10^6$. On the other hand, in the case of acceleration at a collisionless shock, explored here, and also considered by Kennel \& Coroniti (1984), a small population of particles absorbs a large fraction of the upstream kinetic energy, leading to a maximum Lorentz factor that is far higher than the value obtained in the case of a collisional shock with heating rather than acceleration. The channeling of a large fraction of the upstream kinetic energy into a small fraction of the particles is of course a well known characteristic of Fermi acceleration at a shock.

\subsection{Electrostatic Acceleration}
\label{sec:ElecAcc}

In addition to Fermi acceleration, electrons in the vicinity of the pulsar-wind termination shock may also experience significant electrostatic acceleration (Abdo et al. 2011). In principle, electrostatic acceleration can boost the electron energy to a level significantly exceeding that possible when only shock and stochastic acceleration are considered. Hence, electrostatic acceleration has been explored by a number of authors as a natural explanation for the $\gamma$-ray synchrotron emission observed during the Crab nebula flares (e.g., Cerutti et al. 2012; Montani \& Bernardini 2014). However, one must keep in mind that even in the case of electrostatic acceleration, the maximum electron energy is still constrained by the effect of synchrotron burnoff (radiation reaction). It is therefore interesting to calculate the maximum theoretical energy that the electron can achieve under the influence of electrostatic acceleration and synchrotron losses.

Electrostatic acceleration in an electric field of strength $E$, generated in the magnetic reconnection region around the shock, results in a constant momentum gain rate given by
\begin{equation}
<\!\dot p\!>_{\rm elec} = q E \ .
\label{eq21newnew}
\end{equation}
Neglecting all other processes, the electrostatic gain rate, $<\!\dot p\!>_{\rm elec}$, balances the synchrotron loss rate, $<\!\dot p\!>_{\rm sh}$, at the electrostatic equilibrium momentum, $p_{\rm elec}=m_e c \, \gamma_{\rm elec}$, such that
\begin{equation}
\Big(<\!\dot p\!>_{\rm elec} + <\!\dot p\!>_{\rm syn}\Big)
\bigg|_{p=p_{\rm elec}} = 0 \ .
\label{eqElecEquil1}
\end{equation}
We can combine Equations~(\ref{eq18newnew}) and (\ref{eq21newnew}) to obtain
\begin{equation}
\Big(q E - \frac{\sigma_{\rm T} B^2 \gamma^2}{6 \pi}\Big)
\bigg|_{\gamma=\gamma_{\rm elec}} = 0 \ ,
\label{eqElecEquil2}
\end{equation}
which yields for the theoretical electrostatic equilibrium Lorentz factor
\begin{equation}
\gamma_{\rm elec} = \frac{p_{\rm elec}}{m_e c}
= \sqrt{\frac{6 \pi q E}{\sigma_{\rm T} B^2}}
\ ,
\label{eqElecEquil3}
\end{equation}
or, equivalently,
\begin{equation}
\gamma_{\rm elec} = 8.25 \times 10^ 9 \left(\dfrac{B}{200 \, \mu {\rm G}}\right)^{-1/2}
\left(\dfrac{E}{B}\right)^{1/2}
\ .
\label{eqElecEquil3b}
\end{equation}

\subsection{First-Order Gain Rate}
\label{FOGain}

Equations~(\ref{eqpdot}) and (\ref{eq21newnew}) indicate that the momentum gain rates $<\!\dot p\!>_{\rm elec}$ and $<\!\dot p\!>_{\rm sh}$, representing electrostatic and shock acceleration, respectively, are each equal to constant values. Hence the first-order momentum gain rate, $<\!\dot p\!>_{\rm gain}$, appearing in the transport equation~(\ref{eq10newnew}) can be written as the sum
\begin{equation}
<\!\dot p\!>_{\rm gain} = <\!\dot p\!>_{\rm elec} + <\!\dot p\!>_{\rm sh}
\ ,
\label{eqFOFtotal}
\end{equation}
or, equivalently,
\begin{equation}
<\!\dot p\!>_{\rm gain} = q E + q B \xi
= A_0 \, m_e c \ ,
\label{eqFOFtotal2}
\end{equation}
where the first-order acceleration rate constant, $A_0 \propto \rm s^{-1}$, is defined by
\begin{equation}
A_0 \equiv A_{\rm sh} + A_{\rm elec} \ ,
\label{eqA0}
\end{equation}
and the individual shock and electrostatic components, respectively are denoted by
\begin{equation}
A_{\rm sh} \equiv \frac{q B \xi}{m_e c}
= 3517 \ {\rm s}^{-1} \left(\frac{B}{200\,\mu{\rm G}}\right) \, \xi
\label{eqA0shock}
\ ,
\end{equation}
and
\begin{equation}
A_{\rm elec} \equiv \frac{q E}{m_e c}
= 3517 \ {\rm s}^{-1} \left(\frac{E}{B}\right) \left(\frac{B}{200\,\mu{\rm G}}\right)
\label{eqA0elec}
\ .
\end{equation}

\subsection{Particle Escape}
\label{sec:pe}

The one-zone model considered here represents an average over the acceleration and emission regions, and therefore the spatial diffusion of the particles through the nebula is treated implicitly using an escape-probability formalism. In this scenario, the electrons remain in the acceleration region for a mean time $t_{\rm esc}$ before escaping. In order for the model to accurately reflect the geometry of the pulsar-wind environment, the energy dependence of $t_{\rm esc}$ needs to be carefully considered, so that the dominant spatial transport processes on large and small scales are properly treated.

The nature of an electron's propagation through the pulsar wind nebula depends on its momentum, $p$, as depicted in Figure~1. For electrons with small momenta, the Larmor radius, $r_{\rm L}=pc/(qB)$, is much smaller than the pulsar wind termination shock radius, $r_t=10^{17}\,$cm. In this case, the electrons are ``trapped'' in the flow, and the escape of the electrons from the acceleration region is accomplished via advection in the outward direction (e.g., Becker \& Begelman 1986). This advective process is also called shock-regulated escape (SRE; see Steinacker \& Schlickeiser 1989). Conversely, for electrons with large momenta, so that $r_{\rm L} \sim r_t$, the escape occurs via spatial diffusion, with a mean-free path, $\ell$, that is comparable to the Larmor radius $r_{\rm L}$ (see Equation~(\ref{eqMFP})). This is called ``Bohm diffusive escape'' (e.g., Dermer \& Menon 2009). Electrons with an intermediate momentum have a mean-free path that is large enough to diffuse back across the shock into the upstream region (so that they experience additional acceleration), but not so large that they escape from the nebula.

In a proper three-dimensional numerical particle transport model, the large- and small-scale behavior of the spatial diffusion and advection processes would automatically be taken into consideration as part of the simulation. However, since we are employing a simplified one-zone model here, representing an average over the acceleration/radiation regions, we must approximate the correct transport behavior by using a suitable expression for the dependence of the escape timescale $t_{\rm esc}(p)$ on the particle momentum $p$, taking into account both the large- and small-scale behaviors. The remainder of this section focuses on the derivation of the correct functional form for $t_{\rm esc}(p)$.

\subsubsection{Shock-Regulated Escape}

In the shock-regulated escape model, the mean-free path is equal to the Larmor radius, corresponding to the limit of Bohm diffusion, and the electron escape timescale, $t_{\rm esc}$, is proportional to the cycle timescale, $t_{\rm cyc}=r_{\rm L}/c$ (see Equation~(\ref{eq23newnew})). It follows that particles with small momentum $p$ are more likely to be swept away from the shock by advection, rather than scattering back to the upstream side of the shock. Conversely, particles with large momentum have large mean-free paths, and therefore they have a better chance of recycling through the shock and experiencing further acceleration. This advective escape process is referred to as ``shock-regulated escape'' (Steinacker \& Schlickeiser 1989). Based on Equations (\ref{eq14newnew}) and (\ref{eq23newnew}), we can write the momentum dependence of the shock-regulated escape timescale, $t_{\rm SRE}$, as (Jokipii 1987; Gallant \& Achterberg 1999)
\begin{equation}
t_{\rm SRE}(p) = w \frac{r_{\rm L}}{c} \propto p \ ,
\label{eq30}
\end{equation}
where $w$ is a dimensionless constant of order unity that accounts for time dilation and obliquity in the relativistic shock. We can also express Equation~(\ref{eq30}) in the equivalent form
\begin{equation}
t_{\rm SRE}(p) = \frac{p}{C_0 m_e c}
\ ,
\label{eq31}
\end{equation}
where the rate constant for the shock-regulated escape process, $C_0 \propto \rm s^{-1}$, is defined by
\begin{equation}
C_0 \equiv \frac{q B}{w m_e c} = 3517 \ {\rm s}^{-1} \left(\frac{B}{200\,\mu{\rm G}}\right) \, w^{-1}
\ .
\label{eq31b}
\end{equation}
The momentum dependence of $t_{\rm SRE}$ confirms that electrons with small momenta tend to escape more rapidly into the downstream region (via advection) than particles with high momenta, as expected (see Figure~1). The shock-regulated escape process therefore tends to harden the particle distribution, which enhances the high-energy component of the resulting synchrotron spectrum. In Section~\ref{sec:pwsol}, we demonstrate that in the case of pure shock/electrostatic acceleration, the escape rate parameter $C_0$ defined in Equation~(\ref{eq31b}) is linked with the acceleration rate parameter $A_0$ defined in Equation~(\ref{eqA0}) via the relation $m_{\rm sh}=-C_0/A_0$, where $m_{\rm sh}$ is the power-law spectral index of the electron number distribution for the case of pure electrostatic/shock acceleration.

\subsubsection{Bohm Diffusive Escape}

For electrons with very high momentum $p$, the Larmor radius becomes so large that another escape channel becomes available to the electrons, in addition to shock-regulated escape. The additional process is the escape of electrons from the pulsar wind nebula itself, which occurs when the Larmor radius becomes comparable to termination shock radius $r_t$. The mean timescale for ultrarelativistic particles to escape into the outer region of the nebula (beyond the termination shock radius $r_t$) via this process, denoted by $t_{\rm Bohm}$, is therefore a function of the particle momentum, $p$, given by
\begin{equation}
t_{\rm Bohm}(p) \equiv \frac{r_t}{v_{\rm diff}} \ , \qquad
v_{\rm diff} = \frac{c}{r_t/\ell}
\ ,
\label{eq32}
\end{equation}
where $v_{\rm diff}$ is the Bohm diffusion velocity and the mean-free path $\ell$ is given by Equation~(\ref{eqMFP}). Combining relations, we find that
\begin{equation}
t_{\rm Bohm}(p) = \frac{r_t^2 qB}{\eta c^2 p} \equiv \frac{m_e c}{F_0 p}
\ ,
\label{eq33}
\end{equation}
where $F_0 \propto \rm s^{-1}$ is the rate constant for Bohm diffusive escape, defined by
\begin{equation}
F_0 \equiv \frac{\eta m_e c^3}{r_t^2 qB}
= 2.56 \times 10^{-17} \ {\rm s}^{-1} \ \eta
\left(\frac{r_t}{10^{17}\,\rm cm}\right)^{-2}
\left(\frac{B}{200\,\mu{\rm G}}\right)^{-1}
\ .
\label{eq34}
\end{equation}
We can also relate the value of $F_0$ to the synchrotron loss rate constant $B_0$ (Equation~(\ref{eq19newnew})), obtaining
\begin{equation}
\frac{F_0}{B_0} = \frac{6 \pi \eta m_e^2 c^4}{r_t^2 \sigma_{\rm T} q B^3}
= 0.494 \ \eta \left(\frac{r_t}{10^{17}\,\rm cm}\right)^{-2}
\left(\frac{B}{200\,\mu{\rm G}}\right)^{-3}
\ .
\label{eq34b}
\end{equation}
This result demonstrates that in the pulsar-wind application, $F_0 \sim B_0$, which will be useful for constraining the model free parameters when we apply it to the interpretation of the Crab nebula flares in Section~\ref{sec:crabapp}.

Equation~(\ref{eq33}) indicates that electrons with large momentum $p$ have a very small timescale for diffusive escape from the pulsar wind nebula. Of course, the escape timescale cannot be less than the light-crossing time for the termination shock radius, $r_t$, which corresponds to setting the diffusion velocity $v_{\rm diff} = c$, or, equivalently, setting the mean-free path $\ell=r_t$ (Hillas 1984). We can use this idea to estimate the Hillas upper limit, $\gamma_{_{\rm H}}$, for the Lorentz factor of the electrons accelerated in the nebula. By combining Equations~(\ref{eq14newnew}) and (\ref{eqMFP}), we find that the maximum Lorentz factor is given by
\begin{equation}
\gamma_{_{\rm H}} = \frac{r_t q B}{\eta m_e c^2}
= 1.17 \times 10^{10} \ \eta^{-1} \left(\frac{r_t}{10^{17}\,\rm cm}\right)
\left(\frac{B}{200\,\mu{\rm G}}\right)
\ .
\label{eq35}
\end{equation}
For the magnetic field assumed here, $B = 200\,\mu$G, and the termination shock radius $r_t =10^{17}\,$cm, we find that the limiting Lorentz factor is $\gamma_{_{\rm H}} \sim 10^{10}$. The associated maximum synchrotron energy computed by substituting $\gamma_{_{\rm H}}$ into Equation~(\ref{eq1newnew}) is in the GeV range, in agreement with the observed $\gamma$-ray emission from the Crab nebula.

\subsubsection{Net Escape Rate}
\label{sec:netesc}

Taking into consideration Equations~(\ref{eq31}) and (\ref{eq33}), we see that particles in the vicinity of the pulsar wind termination shock have two avenues available for escape from the acceleration region. Particles with small momentum $p$ are likely to advect away into the downstream region, since the shock-regulated escape timescale, $t_{\rm SRE}$, is small in this case according to Equation~(\ref{eq31}). On the other hand, particles with large momentum are likely to rapidly diffuse out of the nebula via Bohm diffusion, since in this case the Bohm diffusion timescale, $t_{\rm Bohm}$, is small, according to Equation~(\ref{eq33}). These two expressions can be combined to write down an expression for the total escape rate, given by
\begin{equation}
t_{\rm esc}^{-1}(p) = t_{\rm SRE}^{-1}(p) + t_{\rm Bohm}^{-1}(p) \ ,
\label{eq36}
\end{equation}
where $t_{\rm esc}(p)$ is the ``effective'' escape timescale, taking both mechanisms into account.

By combining Equations~(\ref{eq31}), (\ref{eq33}), and (\ref{eq36}), we find that $t_{\rm esc}(p)$ can be written as
\begin{equation}
t_{\rm esc}(p) = \left(\frac{C_0 m_e c}{p} + \frac{F_0 p}{m_e c}\right)^{-1} \ .
\label{eq37}
\end{equation}
This is the form for the escape timescale that will be substituted into the transport equation~(\ref{eq10newnew}) in Section~\ref{sec:sste} in order to ensure that both the large- and small-scale behaviors are properly accounted for. The effective escape timescale $t_{\rm esc}(p)$ vanishes in the limits $p \to 0$ and $p \to \infty$, and it is maximized for the momentum value $p_c$, where
\begin{equation}
p_c = \gamma_c \, m_e c = m_e c \, \sqrt{\frac{C_0}{F_0}}
\ .
\label{eq38}
\end{equation}
This is also the ``cross-over'' momentum, at which the individual escape timescales $t_{\rm SRE}$ and $t_{\rm Bohm}$ are equal. For Lorentz factors $\gamma > \gamma_c$, the particle escape is dominated by spatial diffusion, and for $\gamma < \gamma_c$, the escape is dominated by shock-regulated advection.
In our numerical calculations, discussed in Section~\ref{sec:crabapp}, we find that $\gamma_c \sim 10^{10}-10^{11}$, which yields a Larmor radius comparable to the termination-shock radius, $r_t = 10^{17}\,$cm.

\section{Steady-State Transport Equation}
\label{sec:sste}

The synchrotron lifetime given by Equation~(\ref{eq6newnew}) provides a rough estimate for the time it takes the electron distribution to reach equilibrium. The fact that the synchrotron timescale is comparable to the flare duration suggests that the particle distribution during the peak of the flare is close to equilibrium. In this case, we are justified in setting the time derivative in Equation~(\ref{eq10newnew}) equal to zero, and solving the steady-state transport equation. Hence the steady-state particle distribution we will obtain is best interpreted as the electron distribution during the peak of the flare.

\subsection{Green's Function}
\label{sec:Green}

Since Equation~(\ref{eq10newnew}) is linear, it is sufficient to determine the steady-state Green's function, $\green(p,p_0)$, resulting from the reprocessing of monoenergetic seed particles, with source term
\begin{equation}
\dot f_{\rm source}(p) = \frac{\dot N_0 \, \delta(p-p_0)}{4 \pi p_0^2} \ ,
\label{eq39}
\end{equation}
corresponding to the continual injection of $\dot N_0$ electrons per unit time with momentum $p_0$.
Once the solution for $\green(p,p_0)$ is known, the steady-state particular solution, $f(p)$, corresponding to an arbitrary source term, $\dot f_{\rm source}(p)$, can be obtained using the convolution
\begin{equation}
f(p) = \int_0^\infty \frac{4 \pi p_0^2}{\dot N_0} \,
\green(p,p_0) \, \dot f_{\rm source}(p_0) \, dp_0
\ .
\label{eq40}
\end{equation}
By combining Equations~(\ref{eq10newnew}), (\ref{eq17newnew}), (\ref{eq18newnew}), (\ref{eqFOFtotal2}), (\ref{eq37}), and (\ref{eq39}), we find that in the pulsar wind nebula, the fundamental steady-state transport equation given by
\begin{align}
\frac{\partial f}{\partial t} = 0 = - \frac{1}{p^2}\frac{\partial}
{\partial p}\bigg[p^2\bigg(-D_0 m_e c \, p\frac{\partial f}{\partial p}
&+ A_0 m_e c f - \frac{B_0 p^2}{m_e c} f \bigg)\bigg] \nonumber \\
&- \bigg(\frac{C_0 m_e c}{p}+\frac{F_0 p}{m_e c}\bigg) f
+ \frac{\dot N_0 \, \delta(p-p_0)}{4 \pi p_0^2} \ ,
\label{eq41}
\end{align}
where $p_0$ is the momentum of the injected electrons and $\dot N_0$ is the injection rate.

It is convenient to transform from the variables $(p,t)$ to the dimensionless momentum, $x$, and the dimensionless time, $y$, defined by
\begin{equation}
x \equiv \frac{p}{m_e c}, \qquad
x_0 \equiv \frac{p_0}{m_e c}, \qquad
y \equiv D_0 t
\ .
\label{eq42}
\end{equation}
In general, the relationship between $x$ and the Lorentz factor $\gamma$ is given by $x = \sqrt{\gamma^2-1}$. Hence, for the ultrarelativistic ($x \gg 1$) electrons responsible for creating the $\gamma$-rays from the Crab pulsar wind nebula, we can write $x=\gamma$ without making any significant error. In terms of the new coordinates $(x,y)$, the steady-state transport equation can be written as
\begin{equation}
\frac{\partial \green}{\partial y} = 0 = \frac{1}{x^2}\frac{\partial}
{\partial x}\left[x^2\left(x\frac{\partial \green}{\partial x}
- \Atilde \green + \Btilde x^2\green\right)\right]
- \frac{\Ctilde \green}{x} - \Ftilde x \green
+ \frac{\dot N_0 \, \delta(x-x_0)}{4 \pi D_0 (m_e c)^3 x_0^2} \ ,
\label{eq43}
\end{equation}
where we have defined the dimensionless constants $\Atilde$, $\Btilde$, $\Ctilde$, and $\Ftilde$, using
\begin{equation}
\Atilde \equiv \frac{A_0}{D_0}, \qquad \Btilde \equiv
\frac{B_0}{D_0}, \qquad \Ctilde \equiv \frac{C_0}{D_0}
\qquad \Ftilde \equiv \frac{F_0}{D_0}
\label{eq44}
\ .
\end{equation}

\subsection{Fokker-Planck Equation}

It is instructive to rewrite Equation~(\ref{eq43}) in the form of a Fokker-Planck equation by defining the electron number distribution, $\Ngreen$, using
\begin{equation}
\Ngreen(x,x_0,y) \equiv 4 \pi (m_e c)^3 x^2 \green(x,x_0,y)
\label{eq45}
\ .
\end{equation}
Note that the total number of electrons, $N_e$, is related to $\Ngreen$ via (cf. Equation~(\ref{eq11newnew}))
\begin{equation}
N_e(y) = \int_0^\infty \Ngreen(x,x_0,y) \, dx
\label{eq46}
\ .
\end{equation}
Using Equation~(\ref{eq45}) to substitute for $\green$ in Equation~(\ref{eq43}) and rearranging terms yields the steady-state Fokker-Planck equation,
\begin{equation}
\frac{\partial \Ngreen}{\partial y} = \frac{\partial^2}{\partial x^2}
\left(\frac{1}{2} \, \frac{d\sigma^2}{dy} \, \Ngreen\right)
- \frac{\partial}{\partial x}\left(\Big<\frac{dx}{dy}\Big>
\, \Ngreen\right) - \frac{\Ctilde}{x} \, \Ngreen - \Ftilde x \Ngreen
+ \frac{\dot N_0 \, \delta(x-x_0)}{D_0} = 0 \ ,
\label{eq47}
\end{equation}
where the ``broadening'' and ``drift'' coefficients are given, respectively, by
\begin{equation}
\frac{1}{2} \, \frac{d\sigma^2}{dy} = x \ , \ \ \ \ \ 
\Big<\frac{dx}{dy}\Big> = 3 + \Atilde - \Btilde \, x^2
\ .
\label{eq48}
\end{equation}
These expressions for the Fokker-Planck coefficients will be used in Section~\ref{sec:EnergyBudget} to help us analyze the energy budget of the observed flares, and to determine the relative importance of electrostatic and stochastic acceleration in creating the distribution of relativistic electrons.

\section{PARTICLE DISTRIBUTION AND RADIATION SPECTRUM}
\label{sec:pdars}

In this section, we obtain the closed-form solution for the electron Green's function, $\Ngreen$, representing the number distribution of electrons in dimensionless momentum $x$ space (or equivalently, in Lorentz factor space $\gamma$). Since the synchrotron lifetime is comparable to the flare duration, we will assume that the electrons reach an approximate equilibrium during the peak of the flare, and we will therefore employ a steady-state transport equation. Once we obtain the solution for the electron number distribution, we will convolve it with the synchrotron emission function to obtain the $\gamma$-ray spectrum emitted by the relativistic electrons accelerated at the pulsar wind termination shock.

\subsection{Electron Green's Function}

In a steady-state situation, $\Ngreen$ satisfies the ordinary differential equation
\begin{equation}
x \, \frac{d^2 \Ngreen}{dx^2} + (\Btilde \, x^2 - 1 - \Atilde) \,
\frac{d \Ngreen}{dx} + \Big(2 \Btilde \, x
- \frac{\Ctilde}{x} - \Ftilde x\Big) \, \Ngreen
= - \frac{\dot N_0 \, \delta(x-x_0)}{D_0}
\ .
\label{eq49}
\end{equation}
The Green's function, $\Ngreen$, must be continuous at the injection momentum, $x=x_0$, and its derivative displays a jump there, which can be evaluated by integrating Equation~(\ref{eq49}) with respect to $x$ over a small region surrounding $x_0$. The result obtained is
\begin{equation}
\lim_{\delta\to 0} \frac{d\Ngreen}{dx}\Bigg|_{x_0+\delta}
- \frac{d\Ngreen}{dx}\Bigg|_{x_0-\delta}
= - \frac{\dot N_0}{D_0 x_0}
\ .
\label{eq50}
\end{equation}
The fundamental solutions to the homogeneous equation obtained when $x \ne x_0$, satisfying appropriate boundary conditions at large and small values of $x$, can be expressed in terms of the Whittaker functions $M_{\kappa,\mu}$ and $W_{\kappa,\mu}$ using
\begin{eqnarray}
\Ngreen(x,x_0) \ \propto \ e^{-\Btilde x^2/4} \ x^{\Atilde/2}
\begin{cases}
M_{\kappa,\mu}(\Btilde x^2/2) \ , & x \le x_0 \ , \\
W_{\kappa,\mu}(\Btilde x^2/2) \ , & x \ge x_0 \ ,
\end{cases}
\label{eq51}
\end{eqnarray}
where the parameters $\kappa$ and $\mu$ are defined by
\begin{equation}
\kappa \equiv 1 + \frac{\Atilde}{4} - \frac{\Ftilde}{2 \Btilde} \ , \qquad
\mu \equiv \frac{\sqrt{(2+\Atilde)^2 + 4 \, \Ctilde}}{4}
\ .
\label{eq52}
\end{equation}

The continuity of the Green's function at $x=x_0$ implies that we can express the global solution for $\Ngreen$ using
\begin{equation}
\Ngreen(x,x_0) = Q_0
\left(\frac{x}{x_0}\right)^{\Atilde/2} e^{-\Btilde(x^2-x_0^2)/4}
M_{\kappa,\mu}\bigg(\frac{\Btilde x^2_{\rm min}}{2}\bigg)
W_{\kappa,\mu}\bigg(\frac{\Btilde x^2_{\rm max}}{2}\bigg)
\label{eq53}
\ ,
\end{equation}
where the normalization constant $Q_0$ is determined by applying the derivative jump condition, and we have made the definitions
\begin{equation}
x_{\rm min} \equiv \min(x,x_0) \ , \qquad
x_{\rm max} \equiv \max(x,x_0) \ .
\label{eq54}
\end{equation}
Substituting Equation~(\ref{eq53}) into Equation~(\ref{eq50}) yields
\begin{equation}
\Btilde x_0 Q_0 \bigg[
M_{\kappa,\mu}\bigg(\frac{\Btilde x_0^2}{2}\bigg)
W'_{\kappa,\mu}\bigg(\frac{\Btilde x_0^2}{2}\bigg)
- W_{\kappa,\mu}\bigg(\frac{\Btilde x_0^2}{2}\bigg)
M'_{\kappa,\mu}\bigg(\frac{\Btilde x_0^2}{2}\bigg)\bigg]
= - \frac{\dot N_0}{D_0 x_0}
\label{eq55}
\ .
\end{equation}
We can evaluate the Wronskian in the square brackets using (Abramowitz \& Stegun 1970)
\begin{equation}
M_{\kappa,\mu}(z)W'_{\kappa,\mu}(z)
- W_{\kappa,\mu}(z)M'_{\kappa,\mu}(z)
= - \frac{\Gamma(1+2\mu)}{\Gamma(\mu-\kappa+1/2)} \ .
\label{eq56}
\end{equation}
Combining Equations~(\ref{eq55}) and (\ref{eq56}), we obtain for the normalization coefficient
\begin{equation}
Q_0 = \frac{\dot N_0\Gamma{(\mu-\kappa+1/2)}}{\Btilde
D_0\Gamma{(1+2\mu)}x_0^2}
\ ,
\label{eq57}
\end{equation}
which can be substituted into Equation~(\ref{eq53}) to obtain the final result for the electron Green's function,
\begin{equation}
\Ngreen(x,x_0) = \frac{\dot N_0\Gamma{(\mu-\kappa+1/2)}}
{\Btilde D_0\Gamma{(1+2\mu)} \, x_0^2}
\left(\frac{x}{x_0}\right)
^{\Atilde/2}e^{-\Btilde(x^2-x_0^2)/4}
M_{\kappa,\mu}\bigg(\frac{\Btilde x_{\rm min}^2}{2}\bigg)
W_{\kappa,\mu}\bigg(\frac{\Btilde x_{\rm max}^2}{2}\bigg)
\label{eq58}
\ ,
\end{equation}
where $\kappa$ and $\mu$ are given by Equations~(\ref{eq52}) and $x_{\rm min}$ and $x_{\rm max}$ are given by Equations~(\ref{eq54}). The solution to the steady-state transport equation given by Equation~(\ref{eq58}) represents the electron distribution resulting from a balance between particle injection, acceleration, energy losses, and particle escape.

The electron distribution given by Equation~(\ref{eq58}) can be used to compute the theoretical synchrotron spectrum produced from a population of radiating relativistic electrons accelerated in the nebula under the combined action of stochastic MHD wave-particle interactions, electrostatic acceleration, shock acceleration, particle escape, and synchrotron losses. The distributions achieved during the peak of each of the observed Crab nebula $\gamma$-ray fares are plotted and discussed in Section~\ref{sec:crabapp}. Each distribution displays a cusp centered at the injection momentum, $x_0$, surrounded by power-law wings. On the high-energy side, the distribution terminates in an exponential cutoff, where synchrotron losses overwhelm particle acceleration. In Section~\ref{sec:crabapp}, we will compare the model predictions with the observational $\gamma$-ray data and analyze the energetics of the flares.

\subsection{Approximate Power-Law Solution for $\Ngreen$}
\label{sec:apppls}

Equation~(\ref{eq58}) for the electron number distribution, $\Ngreen$, represents the exact solution to the steady-state Fokker-Planck equation~(\ref{eq47}). It is interesting to note that for values of the particle momentum $p$ far below the onset of the synchrotron losses in Equation~(\ref{eq47}), and also below the cross-over momentum indicating the onset of Bohm diffusion (see Equation~(\ref{eq38})), we find that Equation~(\ref{eq47}) reduces to an equidimensional equation, which implies the existence of
power-law solutions of the form
\begin{equation}
\Ngreen(x,x_0)=H_0 x^m
\label{eq59}
\ ,
\end{equation}
where $H_0$ is a normalization constant and $m$ is an unknown power-law index. By substituting the power-law form $\Ngreen(x) \propto x^m$ into Equation~(\ref{eq49}) and simplifying, we can obtain a quadratic equation for $m$, given by
\begin{equation}
m^2 - (2+\tilde{A}) \, m - \tilde{C} = 0
\ ,
\label{eq60}
\end{equation}
with corresponding solutions
\begin{equation}
m_{\pm} = \frac{2+\Atilde \pm \sqrt{(2+\Atilde)^2+4 \Ctilde}}{2}
\label{eq61}
\ .
\end{equation}
Here, the positive power-law index $m_+$ applies at low energies ($x < x_0$), and the negative index $m_-$ applies at high energies ($x > x_0$). The global solution for $\Ngreen$ can now be written as
\begin{equation}
\Ngreen(x,x_0) = \ H_0
\begin{cases}
\Big(\frac{x}{x_0}\Big)^{m_+} \ , & x \le x_0 \ , \\
\Big(\frac{x}{x_0}\Big)^{m_-} \ , & x \ge x_0 \ ,
\end{cases}
\label{eq62}
\end{equation}
where the normalization constant $H_0$ can be determined via application of the derivative jump condition given by Equation~(\ref{eq50}). After some algebra, the result obtained is
\begin{equation}
H_0=\frac{\dot{N}_0}{4D_0 \mu}
\label{eq63}
\ .
\end{equation}
This normalization coefficient can be substituted back into Equation~(\ref{eq62}) to obtained the properly normalized global solution for $\Ngreen$, given by
\begin{equation}
\Ngreen(x,x_0) = \ \frac{\dot{N}_0}{4D_0 \mu}
\begin{cases}
\Big(\frac{x}{x_0}\Big)^{m_+} \ , & x \le x_0 \ , \\
\Big(\frac{x}{x_0}\Big)^{m_-} \ , & x \ge x_0 \ 
\end{cases}
\label{eq64}
\end{equation}
The broken power-law solution given by Equation~(\ref{eq64}) is valid if we restrict attention to values of $x$ below the exponential turnover created by synchrotron losses, and also below the cross-over Lorentz factor, $\gamma_c$, where the transition to Bohm diffusive escape occurs (see Equation~(\ref{eq38})). We will compare the approximate power-law solution with the exact solution in our applications to the Crab nebula flares in Section~\ref{sec:crabapp}.

\subsection{Power-Law Index for Electrostatic/Shock Acceleration}
\label{sec:pwsol}

We have demonstrated that for energies below the onset of synchrotron losses, the particle distribution is well represented by a broken power-law. A case of particular interest is the case of pure electrostatic/shock acceleration, which corresponds to the limit $D_0 \to 0$, where $D_0$ is the momentum diffusion rate coefficient. Physically, momentum diffusion is the result of stochastic wave-particle interactions. In the limit $D_0 \to 0$, the contribution to the acceleration due to the random motions of the MHD waves vanishes, and we are left with only the contribution due to electrostatic/shock acceleration. We can explore this limit in detail by using Equations~(\ref{eq44}) to make the substitutions $\Atilde =A_0/D_0$ and $\Ctilde=C_0/D_0$ in Equation~(\ref{eq61}) for the power-law index $m_\pm$. The result obtained for the high-energy index, $m_-$, is
\begin{equation}
m_- = \Big(1+\frac{A_0}{2D_0}\Big) - \frac{1}{2}\sqrt{\Big(2+\frac{A_0}{D_0}\Big)^2
+ \frac{4C_0}{D_0}}
\label{eq65} \ ,
\end{equation}
which can be rewritten as
\begin{equation}
m_- = \Big(1+\frac{A_0}{2D_0}\Big)
\bigg[1 - \sqrt{1+\frac{4D_0 C_0}{A_0^2}\Big(\frac{2D_0}{A_0}+1\Big)^{-2}}\bigg]
\label{eq66}
\ .
\end{equation}
Making an expansion in terms of the small parameter $D_0/A_0$ and keeping only the highest-order term yields
the power-law index of the electron number distribution for the case of pure electrostatic/shock acceleration. After some algebra, the result obtained is
\begin{equation}
m_{\rm sh} \equiv \lim_{D_0 \to 0} m_-
= -\frac{C_0}{A_0} = - \frac{\Ctilde}{\Atilde}
\label{eq67}
\ .
\end{equation}
In the case of strong electrostatic/shock acceleration, we expect to find that the high-energy power-law index $m_{\rm sh}$ is in the range $-3 \lapprox m_{\rm sh} \lapprox -2$, as is typically found in PIC simulations of acceleration in regions of magnetic reconnection near the Crab pulsar termination shock (e.g., Cerutti et al. 2014a).

\subsection{Synchrotron Spectrum}

The $\gamma$-rays emitted during the recent flares observed from the Crab nebula may represent direct synchrotron radiation produced by relativistic electrons accelerated at the pulsar wind termination shock (Buehler at al. 2012; Abdo et al. 2011). Alternatively, the emission could be produced at the base of the pulsar jet (Weisskopf et al. 2013), although, to our knowledge, no complete physical model of this type has been developed that can reproduce the observed flare spectra. We will focus on the first possibility here, and in this section we explore the implications of our particle transport model for the production of $\gamma$-ray synchrotron emission in the vicinity of the termination shock.

Since synchrotron losses are included in the transport equation we have solved (Equation~(\ref{eq41})), we are now in a position to self-consistently calculate the resulting $\gamma$-ray spectrum. Assuming an isotropic distribution of electrons, the theoretical synchrotron spectrum can be computed by convolving the electron Green's function (Equation~(\ref{eq58})) with the synchrotron emission function, $P_\nu$, which gives the power emitted per electron per Hz. The isotropic synchrotron emission function is given by (e.g., Rybicki \& Lightman 1979)
\begin{equation}
P_\nu(\nu,\gamma) = \frac{\sqrt{3}\,q^3 B}{m_e c^2}R
\left(\frac{\nu}{\gamma^2 \nu_s}\right) \ \ \propto \ \ {\rm erg \ s^{-1}
\ Hz^{-1}}
\ ,
\label{eq68}
\end{equation}
where
\begin{equation}
\nu_s \equiv \frac{3 q B}{4 \pi m_e c}
\ ,
\label{eq69}
\end{equation}
and (Crusius \& Schlickeiser 1986)
\begin{equation}
R(x) \equiv
\frac{x^2}{2}K_{4/3}\Big(\frac{x}{2}\Big)K_{1/3}\Big(\frac{x}{2}\Big)-
\frac{3x^3}{20}\Big[K^2_{4/3}\Big(\frac{x}{2}\Big)-
K^2_{1/3}\Big(\frac{x}{2}\Big)\Big]
\ .
\label{eq70}
\end{equation}
Here, $K_{4/3}(x)$ and $K_{1/3}(x)$ denote modified Bessel functions of the second kind.
The synchrotron spectrum emitted by the entire electron distribution is computed by performing the integral convolution
\begin{equation}
P_\nu^{\rm tot}(\nu) = \int_1^\infty \Ngreen(\gamma,\gamma_0)
P_\nu(\nu,\gamma)d\gamma \ \propto \ {\rm erg \ s^{-1} \ Hz^{-1}}
\ ,
\label{eq71}
\end{equation}
where $\Ngreen$ is evaluated using the analytic solution for the electron distribution given by Equation~(\ref{eq58}). The corresponding observational flux levels are given by
\begin{equation}
\mathscr{F}_\nu(\nu) = \frac{1}{4 \pi D^2} \int_1^\infty
\Ngreen(\gamma,\gamma_0) P_\nu(\nu,\gamma)d\gamma
\ \propto \ {\rm erg \ s^{-1} \ cm^{-2} \ Hz^{-1}}
\ ,
\label{eq72}
\end{equation}
where $D$ is the distance to the source and $\Ngreen$ is given by Equation~(\ref{eq58}), and we remind the reader that $\gamma=x$ for the ultrarelativistic electrons of interest here.

\section{APPLICATION TO THE CRAB NEBULA FLARES}
\label{sec:crabapp}

In this section, we apply our model for the transport and acceleration of relativistic electrons at the pulsar wind termination shock to attempt to understand the nature of the $\gamma$-ray flares observed from the Crab nebula using {\it Fermi}-LAT. Application of our model requires the specification of the dimensionless theory parameters $\Atilde$, $\Btilde$, $\Ctilde$, and $\Ftilde$, the magnetic field strength, $B$, the Lorentz factor of the injected electrons, $x_0$, and the electron injection rate, $\dot N_0$. We remind the reader that the parameters $\Atilde$, $\Btilde$, $\Ctilde$, and $\Ftilde$ describe, in turn, the effects of electrostatic/shock acceleration, synchrotron losses, shock-regulated particle escape, and Bohm diffusive particle escape. By combining Equations~(\ref{eq17Bnewnew}), (\ref{eqA0}), (\ref{eqA0shock}), (\ref{eqA0elec}), and (\ref{eq44}), we find that the dimensionless theory parameter $\Atilde$ is related to the electric field $E$, the magnetic field $B$, and the magnetization parameter $\sigma_{\rm mag}$ via
\begin{equation}
\Atilde = \Atilde_{\rm sh} + \Atilde_{\rm elec}
\label{eqPhyProp1}
\ ,
\end{equation}
where
\begin{equation}
\Atilde_{\rm sh} \equiv \frac{A_{\rm sh}}{D_0} = \frac{3 \eta \, \xi}{\sigma_{\rm mag}} \ , \qquad
\Atilde_{\rm elec} \equiv \frac{A_{\rm elec}}{D_0} = \frac{E}{B} \frac{3 \eta}{\sigma_{\rm mag}}
\label{eqPhyProp2}
\ .
\end{equation}
Likewise, we can combine Equations~(\ref{eq17Bnewnew}), (\ref{eq19newnew}), and (\ref{eq44}) to show that the theory parameter $\Btilde$ can be expressed in terms of $B$ and $\sigma_{\rm mag}$ using
\begin{equation}
\Btilde = \frac{\sigma_{\rm T} \eta B}{2 \pi q \sigma_{\rm mag}}
= 4.41 \times 10^{-20} \left(\frac{B}{200\,\mu{\rm G}}\right) \eta \, \sigma_{\rm mag}^{-1}
\label{eqPhyProp3}
\end{equation}
Next, we can combine Equations~(\ref{eq17Bnewnew}), (\ref{eq31b}), and (\ref{eq44}) to write $\Ctilde$ in terms of $\sigma_{\rm mag}$, obtaining
\begin{equation}
\Ctilde = \frac{3 \eta}{w \sigma_{\rm mag}} \ .
\label{eqPhyProp4}
\end{equation}
Finally, we can combine Equations~(\ref{eq17Bnewnew}), (\ref{eq34b}), and (\ref{eq44}) to show that $\Ftilde$ is related to $\Btilde$, $r_t$, and $B$ via
\begin{equation}
\Ftilde = 0.494 \ \eta \left(\frac{r_t}{10^{17}\,\rm cm}\right)^{-2}
\left(\frac{B}{200\,\mu{\rm G}}\right)^{-3} \, \Btilde
\ .
\label{eqPhyProp5}
\end{equation}
In our approach, we treat $\Atilde$, $\Btilde$, and $\Ctilde$ as free parameters, and we compute $\Ftilde$ using Equation~(\ref{eqPhyProp5}). Hence $\Ftilde$ is not a free parameter in our model.

In our consideration of the $\gamma$-ray flares from the Crab nebula, we adopt for the magnetic field strength the value $B=200\,\mu$G, implied by multiple studies of the quiescent emission from the Crab nebula (e.g., Aharonian et al. 2004; Meyer et al. 2010), and the value used for the termination shock radius is $r_t = 10^{17}\,$cm (Montani \& Bernardini 2014). The model free parameters $\Atilde$, $\Btilde$, $\Ctilde$, $x_0$, and $\dot N_0$ are varied until a reasonable qualitative fit to the $\gamma$-ray spectral data is obtained for a given flare. We set $\eta=1$ and $\xi=0.1$ in all of our numerical calculations. Once the model parameter values have been established, the momentum diffusion rate parameter $D_0$ can be evaluated using $\Btilde$ and $B$ by combining Equations~(\ref{eq17Bnewnew}) and (\ref{eq19newnew}) to obtain
\begin{equation}
D_0 = \frac{\sigma_{\rm T} B^2}{6 \pi m_e c \Btilde} \ ,
\label{eq93}
\end{equation}
and the magnetization parameter $\sigma_{\rm mag}$ can then be determined by using Equation~(\ref{eq17Bnewnew}) to write
\begin{equation}
\sigma_{\rm mag} = \frac{3 \eta m_e c D_0}{q B}
\ .
\label{eq93b}
\end{equation}
Once the values of $\Ctilde$ and $\sigma_{\rm mag}$ have been obtained, we can compute the SRE timescale constant $w$ introduced in Equation~(\ref{eq30}) by using Equation~(\ref{eqPhyProp4}) to write
\begin{equation}
w = \frac{3 \eta}{\Ctilde \sigma_{\rm mag}} \ .
\label{eq93d}
\end{equation}
Next, we can obtain the electric field ratio, $E/B$, by combining Equations~(\ref{eqPhyProp1}) and (\ref{eqPhyProp2}), which yields
\begin{equation}
\frac{E}{B} = \frac{\Atilde \, \sigma_{\rm mag}}{3 \eta} - \xi
\ .
\label{eq93c}
\end{equation}

In Figure~2, we plot the $\gamma$-ray spectra computed using Equation~(\ref{eq72}) along with the spectra observed during the peak for each of the five flares observed by {\it Fermi}-LAT and {\it AGILE} and discussed by Abdo et al. (2011), Buehler et al. (2012), Buehler \& Blandford (2014), and Striani et al. (2013). The model parameters were varied in order to obtain a reasonably good qualitative fit to the $\gamma$-ray data for each flare. The theoretical $\gamma$-ray spectra display a power-law shape at low energies, leading up to a broad peak where most of the flare energy is emitted, followed by an exponential decrease at high energies. It is clear from Figure~2 that the analytical electron transport model considered here is able to roughly reproduce the observed $\gamma$-ray spectra for each of the observed {\it Fermi}-LAT flares. We have set $\eta=1$, $\xi=0.1$, and $B=200\,\mu$G in all of the calculations, and the values for all of the model free parameters are reported in Table~1. We also list the values obtained for the cross-over Lorentz factor, $\gamma_c$, (see Equation~(\ref{eq38})) which represents the transition energy between shock-regulated escape at low energies and Bohm diffusive escape at high energies. In our applications to the Crab nebula flares, we find that $\gamma_c \sim 10^{10}-10^{11}$, and therefore $\gamma_c \gapprox \gamma_{_{\rm H}}$, where $\gamma_{_{\rm H}}$ (see Equation~(\ref{eq35})) is the Hillas (1984) limit, above which the electrons are not confined to the nebula. This result implies that Bohm diffusion is not a very important contributor to the escape of particles from the acceleration region, and therefore particle escape is dominated by the shock-regulated escape (SRE) mechanism.

The corresponding electron distributions for each flare are plotted in Figures~3 and 4. The plots include a comparison of the exact solution for $\Ngreen$ computed using Equation~(\ref{eq58}) with the approximate broken power-law solution given by Equation~(\ref{eq64}). We note that the agreement between the approximate and exact solutions is excellent, up to the energy where synchrotron losses become dominant, and the electron distribution transitions into an exponential turnover. The electron distributions for the 2009 February, 2010 September, and 2013 March flares are plotted in Figure~3. In each of these cases, the distribution function closely resembles the broken power-law solution (Equation~(\ref{eq64})), with the break occurring at the injection Lorentz factor $\gamma_0$. On the high-energy side, the spectrum has a power-law shape up to the exponential turnover, at $\gamma \sim 10^{10}$, and there is no particle pile-up. We note that the maximum value of $\gamma$ is in good agreement with the predicted upper limit $\gamma_{_{\rm H}} \sim 10^{10}$ set by the Hillas condition (see Equation~(\ref{eq35})).

The electron distributions for the 2007 September and 2011 April flares are plotted in Figure~4. In these two cases, the distribution functions resemble the approximate broken power-law solution at low energies (Equation~(\ref{eq64})), but they each display a distinctive pile-up around the maximum Lorentz factor, $\gamma \sim 10^{10}$, resulting in the sharply peaked $\gamma$-ray spectra for these two flares, as depicted in Figure~2. The maximum particle energy is in agreement with the Hillas upper limit, $\gamma_{_{\rm H}} \sim 10^{10}$, given by Equation~(\ref{eq35}).

Table~1 also includes the values obtained for the high-energy power-law index, $m_-$, computed using Equation~(\ref{eq61}). The 2007 September and 2011 April flares have the flattest high-energy power-law index for the electron distribution, with $m_- \sim -0.3$, suggesting that extremely efficient electrostatic/shock acceleration is occurring during these two flares. Hence we view the 2007 September and 2011 April flares as indicative of the strongest particle acceleration ever observed in the Crab nebula. We can determine the specific amount of acceleration associated with the shock and the electric field by using Equations~(\ref{eqA0shock}) and (\ref{eqA0elec}) to compute $\Atilde_{\rm sh}$ and $\Atilde_{\rm elec}$, respectively, and the results are listed in Table~1. One can see that electrostatic acceleration dominates in each flare model, as expected (e.g., Cerutti et al. 2012), and substantial electrostatic acceleration is required to explain the $\gamma$-ray flare spectra. The inferred electric field values in the magnetic reconnection layer found using Equation~(\ref{eq93c}) falls in the range $E \sim 50-600\,\mu$G in Gaussian units for an ambient magnetic field $B = 200\,\mu$G. These values satisfy the condition $E \gapprox B$, which is consistent with rapid magnetic reconnection, giving rise to efficient electrostatic acceleration. The only exception is the weakest flare, observed in 2009 February, for which we obtain $E/B\sim 0.2$. However, this particular flare barely exceeded the level of the quiescent nebular emission, so our results are reasonable in the sense that strong electrostatic acceleration is not required to explain the spectrum observed during that flare.

The values for the magnetization parameter $\sigma_{\rm mag}$ obtained by substituting $B$, $\eta$, and $D_0$ into Equation~(\ref{eq93b}) are reported in Table~1. We find that $0.04 \lapprox \sigma_{\rm mag} \lapprox 0.7$, which is within the range deduced by Mori et al. (2004) in their analysis of the asymmetry of the X-ray brightness between the far and near sides of the equatorial region of the nebula. However, it should be emphasized that the value of $\sigma_{\rm mag}$ is not well constrained by the observations or the models, and could range from $\sigma_{\rm mag} \sim 10^{-3}$ in the magnetohydrodynamical models (e.g., Kennel \& Coroniti 1984) up to $\sigma_{\rm mag} \sim 1$ in the striped wind models (e.g., Komissarov 2013).

Table~1 includes the values used in each flare model for the Lorentz factor of the injected electrons, $\gamma_0=x_0$, and for the particle injection rate, $\dot N_0$. The associated power in the injected particles is given by $P_{\rm inj} = \gamma_0 m_e c^2 \dot N_0$, assuming isotropic emission. We confirm in each case that the injected power $P_{\rm inj}$ does not exceed the pulsar spin-down power, which is $\sim 5 \times 10^{38}\,\rm erg\,s^{-1}$. In general, we set $\gamma_0=10^6$ in order to simulate the effect of the injection of electrons from the ``cold'' pulsar wind, in which the electrons have a high bulk Lorentz factor but a small random component (Lyubarsky 2003). However, in the case of the 2013 March flare, we find it necessary to set $\gamma_0=5 \times 10^8$ in order to avoid an injection power $P_{\rm inj}$ that exceeds the spin-down power. The value $\gamma_0=5 \times 10^8$ is much higher than expected for the cold pulsar wind, but it is in the expected range if one considers the absorption of the electromagnetic Poynting flux by the electrons near the termination shock, leading to a ``hot'' input distribution rather than a cold one (Rees \& Gunn 1974). This is essentially the scenario considered by Cerutti et al. (2014b), who assumed that the injected electrons were sampled from an ultrarelativistic Maxwellian distribution with temperature $k T/(m_e c^2)=10^8$. Similarly, Cerutti et al. (2013) assumed the injection of a power-law electron distribution extending up to a maximum Lorentz factor equal to $4 \times 10^8$. We also note that the SRE timescale parameter $w \sim 0.1$ (Equation~(\ref{eq30})) for the 2013 March flare, which is the smallest value for any of the flares. This may reflect a stronger relativistic decrease in the shock cycle time for this flare (Gallant \& Achterberg 1999).

\section{SYNCHROTRON AFTERGLOW}
\label{sec:afterglow}

Although the $\gamma$-ray flares observed from the Crab nebula are intrinsically time-dependent phenomenon, in this paper, we have employed a steady-state approach to model the underlying electron distribution, under the assumption that the electrons reach equilibrium during the peak of the flare. This is reasonable, provided the flare duration timescale is comparable to the synchrotron loss timescale, which is in fact the case, according to Equation~(\ref{eq6newnew}). In our one-zone model, the electrons that create the observed $\gamma$-ray synchrotron flares are accelerated and radiate in the same region, which is in the vicinity of the pulsar wind termination shock.

The electrons that produce the peak level of $\gamma$-ray emission observed during a given flare eventually escape into the downstream (outer) region, at radius $r > r_t$, where $r_t=10^{17}\,$cm is the pulsar wind termination shock radius. The escape of the electrons into the outer region of the nebula occurs via a combination of advection for the low-energy electrons that are ``trapped'' in the outflow, and Bohm diffusion for the high-energy electrons. The transition between these two escape channels occurs at the cross-over Lorentz factor, $\gamma_c \sim 10^{10}-10^{11}$ (see Equation~(\ref{eq38}) and Table~1).

Once the electrons escape, they are subject to continued synchrotron cooling in the outer region of the nebula, but they do not experience any additional acceleration. Since this is a non-equilibrium situation, the synchrotron spectrum emitted by the cooling electrons varies with time, and therefore the electrons in the cooling region produce a variable ``synchrotron afterglow'' spectrum that gradually fades away and shifts to lower frequencies. We can use our model to compute the time-dependent synchrotron afterglow spectrum, and to make predictions that can be compared with multi-wavelength observations of future $\gamma$-ray flares from the Crab nebula.

The transport equation for the escaping electrons in the cooling region is quite simple, since they only experience losses. Here, we will focus solely on the synchrotron losses, in order to obtain an upper limit on the afterglow radiation. If adiabatic losses are also taken into consideration, that will reduce the level of the resulting spectrum below that predicted here. As an escaping electron cools in response to synchrotron losses, its Lorentz factor $\gamma$ varies according to (see Equation~(\ref{eq4newnew}))
\begin{equation}
-\frac{1}{\gamma^2} \frac{d\gamma}{dt} = \frac{\sigma_{\rm T} B_{\rm cool}^2}
{6\pi m_e c}
\ ,
\label{eq96}
\end{equation}
where $B_{\rm cool}$ is the magnetic field in the cooling region. We can rewrite Equation~(\ref{eq96}) as
\begin{equation}
-\dfrac{d\gamma}{\gamma^2}={\mathscr B}_0 dt
\ ,
\label{eq97}
\end{equation}
where the constant ${\mathscr B}_0 \propto \rm s^{-1}$ is given by (cf. Equation~(\ref{eq19newnew}))
\begin{equation}
{\mathscr B}_0 = \frac{\sigma_{\rm T} B_{\rm cool}^2}{6 \pi m_e c} \ .
\label{eq98}
\end{equation}
The solution obtained for the time variation of the Lorentz factor is
\begin{equation}
\gamma(\gamma_*,t) = \left(\dfrac{1}{\gamma_*} + {\mathscr B}_0 t\right)^{-1}
\ ,
\label{eq99}
\end{equation}
where $\gamma_*$ is the initial value of the electron's Lorentz factor at time $t=0$ as it enters the downstream cooling region and begins the cooling phase of its evolution in the nebula. The initial Lorentz factor $\gamma_*$ can be computed in terms of the Lorentz factor $\gamma$ at time $t$ by inverting Equation~(\ref{eq99}) to obtain
\begin{equation}
\gamma_*(\gamma,t) = \left(\frac{1}{\gamma} - {\mathscr B}_0 t\right)^{-1}
\ .
\label{eq100}
\end{equation}
This result implies that the maximum possible Lorentz factor at time $t$ is equal to $1/({\mathscr B}_0 t)$.

The time-dependent electron distribution in the cooling region, denoted by $\Ncool(t,\gamma)$, evolves under the influence of synchrotron losses according to the relation
\begin{equation}
\Ncool(t,\gamma) = J(t) \, \Ncool(0,\gamma_*)
\ ,
\label{eqeq101}
\end{equation}
where $\Ncool(0,\gamma_*)$ is the initial distribution at time $t=0$, and the normalization function $J(t)$ can be determined by requiring that the total number of electrons is conserved during the cooling phase. Conservation of electron number during the cooling phase implies the differential relation
\begin{equation}
\Ncool(t,\gamma) \, d\gamma = \Ncool(0,\gamma_*) \, d\gamma_*
\ ,
\label{eq102}
\end{equation}
or, equivalently,
\begin{equation}
\Ncool(t,\gamma) = \Ncool(0,\gamma_*)
\left(\frac{\partial \gamma_*}{\partial \gamma}\right)_t
\ ,
\label{eq103}
\end{equation}
where $\gamma_*$ is computed using Equation~(\ref{eq100}). The required partial derivative is given by
\begin{equation}
J(t) \equiv \left(\frac{\partial \gamma_*}{\partial \gamma}\right)_t
= \frac{\gamma_*^2}{\gamma^2}
\label{eq104}
\ .
\end{equation}
By combining this result with Equation~(\ref{eqeq101}), we find that the electron distribution in the cooling region after time $t$ is related to the electron distribution at time $t=0$ using
\begin{equation}
\Ncool(t,\gamma) = \gamma^{-2} \left(\frac{1}{\gamma} - {\mathscr B}_0 t\right)^{-2} \,
\Ncool\left[0,\left(\frac{1}{\gamma} - {\mathscr B}_0 t\right)^{-1}\right]
\ ,
\label{eq105}
\end{equation}
where the value of $\gamma$ at time $t$ cannot exceed the limit $\gamma < 1/({\mathscr B}_0 t)$.

Equation~(\ref{eq105}) relates the electron distribution $\Ncool(t,\gamma)$ in the cooling region to the starting distribution $\Ncool(0,\gamma)$ at time $t=0$. Our remaining task is to compute the initial distribution at $t=0$. This can be accomplished by recognizing that the initial distribution in the downstream cooling region is equal to the population of electrons that escapes from the acceleration region. Hence we can write
\begin{equation}
\Ncool(0,\gamma) = t_* \, \Ngreen(\gamma,\gamma_0) \, t_{\rm esc}^{-1}(\gamma)
\ ,
\label{eq106}
\end{equation}
where $\gamma_0$ is the Lorentz factor of the monoenergetic electrons injected into the acceleration region, the acceleration-region particle distribution $\Ngreen$ is computed using Equation~(\ref{eq58}), and $t_*$ is the timescale for electrons to accumulate in the downstream cooling region, before advection sweeps them into the outer part of the nebula.

The advection timescale is independent of the particle energy. In our application to the Crab nebula, we vary $t_*$ such that the afterglow spectrum immediately after a given flare has the same flux level as the peak flare spectrum. The escape timescale, $t_{\rm esc}(\gamma)$, appearing in Equation~(\ref{eq106}) is given by (see Equation~(\ref{eq37}))
\begin{equation}
t_{\rm esc}(\gamma) = \left(\frac{C_0}{\gamma} + F_0 \gamma\right)^{-1} \ ,
\label{eq107}
\end{equation}
which exhibits the variation between shock-regulated escape at low particle energies and Bohm diffusive escape at high particle energies (see Section~\ref{sec:pe}). The transition between the two escape mechanism occurs at the cross-over Lorentz factor, $\gamma_c$, computed using Equation~(\ref{eq38}). In our models, we find that $\gamma_c \sim 10^{10}-10^{11}$ (see Table~1).

By combining Equations~(\ref{eq105}), (\ref{eq106}), and (\ref{eq107}), we find that the advanced-time electron distribution in the cooling region can be written in the explicit form
\begin{equation}
\Ncool(t,\gamma) = \left(\frac{1}{\gamma} - {\mathscr B}_0 t\right)^{-3} \,
t_* \, \gamma^{-2} \,
\left[C_0 \left(\frac{1}{\gamma} - {\mathscr B}_0 t\right)^2
+ F_0 \right]
\Ngreen\left[\left(\frac{1}{\gamma} - {\mathscr B}_0 t\right)^{-1},\gamma_0\right]
\ ,
\label{eq108}
\end{equation}
where $\Ngreen$ is computed using Equation~(\ref{eq58}). The synchrotron afterglow spectrum generated at time $t$ by the population of electrons that have escaped into the downstream (cooling) region is computed using (see Equation~(\ref{eq72}))
\begin{equation}
\mathscr{F}^{\rm cool}_\nu(t,\nu) = \frac{1}{4 \pi D^2} \int_1^{({\mathscr B}_0 t)^{-1}}
\Ncool(t,\gamma) P_\nu(\nu,\gamma)d\gamma
\ \propto \ {\rm erg \ s^{-1} \ cm^{-2} \ Hz^{-1}}
\ ,
\label{eq109}
\end{equation}
where $D$ is the distance to the source, $\Ncool$ is evaluated using Equation~(\ref{eq108}), and $P_\nu(\nu,\gamma)$ is given by Equation~(\ref{eq68}). Note that the upper bound for the integration over $\gamma$ at time $t$ is equal to $({\mathscr B}_0 t)^{-1}$ due to the action of synchrotron cooling.

In Figure~5 we plot the synchrotron afterglow spectra for the each of the $\gamma$-ray flares considered in Figure~2, calculated using Equation~(\ref{eq109}), with the accumulation timescale $t_*$ set so that the initial afterglow flux is comparable to the peak $\gamma$-ray flare spectrum. The spectrum is plotted as a series of ``snapshots,'' corresponding to the elapsed time values $t=1\,$s, $t=9\,$days, and $t=21\,$days. The magnetic field strength in the downstream (cooling) region, $B_{\rm cool}$, is likely to be lower than that in the upstream acceleration region, $B$, but for the calculations in Figure~5, we have set $B_{\rm cool} = B=200\,\mu$G in order to maximize the afterglow. In each case, we find that the afterglow declines to the level of the quiescent emission over a period of 3-4 weeks. However, the levels of emission reported in Figure~5 will be reduced by adiabatic losses in the expanding wind, and also by the decreasing magnetic field strength experienced by the electrons as they advect to larger radii in the wind. We have not included either of these effects in our calculations, and therefore the spectral snapshots plotted in Figure~5 represent upper limits on the predicted level of the afterglow emission.

Afterglow spectra were also computed by Tavani et al. (2011) and by Zrake (2016), assuming synchrotron losses as the only cooling mechanism. They found that the afterglow spectra decay down to quiescent flux levels in a few days, based on an assumed magnetic field strength $B \sim 1000 \, \mu$G. Our results for the afterglow spectra are comparable to theirs, except that ours extend over a longer cooling timescale, which is probably due to the fact that our assumed magnetic field is five times smaller than that adopted in the studies of Tavani et al. (2011) and Zrake (2016).

\section{Flare Energy Budgets}
\label{sec:EnergyBudget}

It is important to analyze the energy budget for each flare by computing the energy loss and gain rates for each term in the transport equation. This will allow us to better visualize the flow of energy from the input side (which includes the energy of the injected particles, plus the energy provided by shock, electrostatic, and MHD acceleration), to the output side (which includes the losses due to synchrotron emission and particle escape). Energy conservation in the model requires that the sum of the input rates must equal the sum of the output rates. We present the formulas used to compute each rate below, and summarize the results obtained for each flare in Table~2.

The ultimate energy source for the $\gamma$-ray flares is the pulsar spin-down power, which arrives on the upstream side of the termination shock primarily in the form of particle kinetic energy, since $\sigma_{\rm mag} \ll 1$ (Kennel \& Coroniti 1984; Begelman 1998). Due to magnetic reconnection and particle acceleration in the vicinity of the pulsar wind termination shock, a large fraction of the kinetic luminosity in the pulsar wind is channeled into a small fraction of particles, which are accelerated up to a maximum Lorentz factor $\gamma \lesssim 10^{10}$.

The total synchrotron power, $P_{\rm syn}$, radiated by the particles is computed by integrating the synchrotron loss rate (Equation~(\ref{eq4newnew})) weighted by the electron number distribution, which yields
\begin{equation}
P_{\rm elec} = \int_1^\infty <\!\dot p\!>_{\rm elec} c N_{\rm G}(\gamma) \, d\gamma
= \int_1^\infty q E c N_{\rm G}(\gamma) \, d\gamma
\ ,
\end{equation}
\begin{equation}
P_{\rm syn} = \int_1^\infty <\!\dot p\!>_{\rm syn} c N_{\rm G}(\gamma) \, d\gamma
= \dfrac{4 \, \sigma_{\rm T} c}{3} \frac{B^2}{8\pi}
\int_1^\infty N_{\rm G}(\gamma) \, \gamma^2 \, d\gamma
\label{eqLstar}
\ ,
\end{equation}
where $N_{\rm G}$ is given by Equation (\ref{eq58}). The flares have an observed (isotropic) luminosity that is about 1\% of the pulsar spin-down power, and our radiated synchrotron power $P_{\rm syn}$ agrees with the observed power, as expected. The synchrotron power for each flare computed using Equation~(\ref{eqLstar}) is listed in Table~2. Another energy loss channel is particle escape, with a timescale that reflects a combination of Bohm diffusion on large scales, and shock-regulated (advective) escape on small scales (see Section~\ref{sec:netesc}). The total power in the escaping particles, $P_{\rm esc}$, is given by (see Equations~(\ref{eq36}) and (\ref{eq37}))
\begin{equation}
P_{\rm esc} = \int_1^\infty(t_{\rm SRE}^{-1}+t_{\rm Bohm}^{-1}) \gamma m_e c^2
N_{\rm G}(\gamma) \, d\gamma
= \int_1^\infty \left(\dfrac{C_0}{\gamma} + F_0\gamma\right) \gamma m_e c^2
N_{\rm G} (\gamma) \, d\gamma
\ ,
\label{eqESC}
\end{equation}
where $C_0$ and $F_0$ are the escape rates for the shock-regulated and Bohm diffusive escape scenarios, respectively. The total energy loss rate for the model is given by the sum
\begin{equation}
P_{\rm loss} = P_{\rm syn} + P_{\rm esc}
\ .
\label{eqPloss}
\end{equation}

Next we consider the energy gain processes in the model. The energy gains due to electrostatic and shock acceleration are represented by the model parameters $\Atilde_{\rm elec}$ and $\Atilde_{\rm sh}$, respectively. The stochastic acceleration due to collisions with the random field of MHD waves is a second-order process and therefore we treat it separately. The total power gained by the particles due to electrostatic acceleration, $P_{\rm elec}$, is computed using (see Equation~(\ref{eq21newnew}))
\begin{equation}
P_{\rm elec} = \int_1^\infty <\!\dot p\!>_{\rm elec} c N_{\rm G}(\gamma) \, d\gamma
= \int_1^\infty q E c N_{\rm G}(\gamma) \, d\gamma
\label{eqELEC}
\ ,
\end{equation}
where $E$ is the electric field. The total input power due to shock acceleration is likewise given by (see Equation~(\ref{eqpdot}))
\begin{equation}
P_{\rm sh} = \int_1^\infty <\!\dot p\!>_{\rm sh} c N_{\rm G}(\gamma) \, d\gamma
= \int_1^\infty \xi B q c N_{\rm G}(\gamma) \, d\gamma
\label{eqSHOCK}
\ ,
\end{equation}
where $\xi$ is the shock acceleration efficiency parameter. The total power provided by stochastic acceleration, $P_{\rm stoch}$, is computed using (see Equation~(\ref{eq75}))
\begin{equation}
P_{\rm stoch} = \int_1^\infty <\!\dot p\!>_{_{\rm MHD}} c N_{\rm G}(\gamma) \, d\gamma
= \int_1^\infty 3 D_0 m_e c^2 N_{\rm G}(\gamma) \, d\gamma
\label{eqMHDgains}
\ ,
\end{equation}
where $D_0$ is given by Equation~(\ref{eq93}). Finally, the power in the injected electrons is given by
\begin{equation}
P_{\rm inj} = \dot N_0 \gamma_0 m_e c^2
\label{eqP0}
\ .
\end{equation}
The total energy gain rate for the model is given by the sum
\begin{equation}
P_{\rm gain} = P_{\rm inj} + P_{\rm elec} + P_{\rm sh} + P_{\rm stoch}
\ .
\label{eqPgain}
\end{equation}
In Table~2 we list the values obtained for all of the gain and loss processes, $P_{\rm syn}$, $P_{\rm esc}$, $P_{\rm elec}$, $P_{\rm sh}$, $P_{\rm stoch}$, and $P_{\rm inj}$, for each of the five flares treated here. It is straightforward to confirm that in each case, $P_{\rm gain} = P_{\rm loss}$, as required for model energy conservation. We find that the $\gamma$-ray flares fall into two main categories, depending on whether most of the energy loss is in the form of $\gamma$-rays, or escaping particles, as discussed below.

\subsection{Efficient Gamma-Ray Flares}

The energy budget figures in Table~2 reveal that the 2011 April and 2007 September flares had the highest efficiency of conversion of input energy into output $\gamma$-ray power. These two flares have high $\Atilde_{\rm elec}$ values and small $\Ctilde$ values, indicating strong confinement in the acceleration region. The attenuation of the particle escape channel means that most of the energy left the system in the form of $\gamma$-ray emission. The low value of $\tilde{C}$ indicates enhanced confinement, possibly reflecting a change in the magnetic topology and/or the obliquity of the termination shock, increasing the probability of additional shock crossings instead of escape. The recycled electrons remained in the vicinity of the termination shock long enough to undergo additional shock and electrostatic acceleration, and the power in the escaping particles for these two flares is two orders of magnitude lower than that for the other flares.

\subsection{Inefficient Gamma-Ray Flares}

The 2009 February, 2010 September, and 2013 March flares all display a pattern in which most of the power leaves the system in the form of escaping particles, rather than $\gamma$-rays. The weakest flare, in 2009 February, displayed flux levels slightly exceeding the quiescent background emission. In this case, the power in the escaping particles exceeds the synchrotron power by two orders of magnitude, and it is also about twice as high as the pulsar spin-down power. This presents a problem, although the factor of two could easily be reduced by supposing either anisotropic emission, or Doppler boosting in the mildly relativistic outflow (Lyutikov et al. 2012). The electron distribution for this flare, plotted in Figure~3, indicates that the peak electron energy is lower than that of all the other flares except the 2013 March transient, which we discuss in further detail below.

The 2013 March flare is unique in that the power in the injected particles, $P_{\rm inj}$, is larger than the power provided by any form of acceleration. Hence, this was essentially a pure synchrotron cooling flare, characterized by a broad, flat-top spectrum that is much less peaked than the other flare spectra (see Figure~2). It is also interesting to note that according to our model, as the afterglow from this flare faded (Figure~5), the emission merged with the well-observed quiescent spectrum after about three weeks. This implies that the electrons escaping into the synchrotron nebula during the 2013 March flare cooled to resemble the quiescent electron distribution. This may be an important clue about the physical mechanism(s) responsible for generating the quiescent electron distribution, which is not understood very well at present (Gaensler \& Slane 2006). The 2013 March flare has the smallest electrostatic gain parameter $\Atilde_{\rm elec}$, but it also has the largest electric field ratio, with $E/B \sim 3$. This apparently paradoxical result is explained by the fact that this flare had the largest magnetization parameter, with $\sigma_{\rm mag} \sim 0.7$, which could imply a unique magnetic field topology in this case.

\section{PARAMETER CONSTRAINTS}
\label{sec:parcon}

In this section we examine the various constraints on the model parameters required in order to ensure that the computational results we have obtained are self-consistent and physically reasonable.

\subsection{MHD Acceleration Limit}
\label{sec:acccon}

Shock acceleration and stochastic acceleration are both mediated by interactions between the electrons and MHD waves, and therefore it follows that the combination of these two processes, operating simultaneously, cannot accelerate electrons beyond the synchrotron burnoff limit, $\gamma_{_{\rm MHD}}$, given by Equation~(\ref{eq8newnew}). Our model must therefore satisfy the condition
\begin{equation}
\Big(<\!\dot p\!>_{\rm sh} + <\!\dot p\!>_{\rm stoch} + <\!\dot p\!>_{\rm syn}\Big)
\bigg|_{p=p_{_{\rm MHD}}} \le 0 \ ,
\label{eq78old}
\end{equation}
where $<\!\dot p\!>_{\rm sh}$ and $<\!\dot p\!>_{\rm stoch}$ denote the mean momentum gain rates for shock acceleration and stochastic acceleration, respectively, and $p_{_{\rm MHD}} \equiv m_e c \, \gamma_{_{\rm MHD}}$. Combining Equation~(\ref{eq78old}) with Equations~(\ref{eq18newnew}), (\ref{eq75}), and (\ref{eqpdot}) yields
\begin{equation}
\Big(\xi q B + 3 D_0 m_e c - \frac{\sigma_{\rm T} B^2 \gamma^2}{6 \pi}\Big)
\bigg|_{\gamma=\gamma_{_{\rm MHD}}} \le 0 \ .
\label{eq110}
\end{equation}
This can be further reduced by using Equation~(\ref{eq8newnew}) to substitute for $\gamma_{_{\rm MHD}}$, which yields
\begin{equation}
\xi + \frac{D_0}{D_0^{\rm max}} \le 1 \ ,
\label{eq118}
\end{equation}
where $D_0^{\rm max}$ is defined by Equation~(\ref{eq80}). This condition must be satisfied in order to ensure that the combination of shock and stochastic acceleration does not lead to particle energies that violate the synchrotron burnoff limit. In Figure~6 we depict the $(\xi,D_0)$ parameter space, with one point plotted for each of the five Crab nebula $\gamma$-ray flares treated here. The results confirm that Equation~(\ref{eq118}) is satisfied for each flare, and therefore our model complies with the synchrotron burnoff limit.

\subsection{Maximum Particle and Radiation Energies}

The transport equation considered here includes shock, electrostatic, and stochastic acceleration, as well as synchrotron losses and particle escape. We can estimate the maximum particle energy achieved in our model by examining the Fokker-Planck ``drift'' coefficient given by Equation~(\ref{eq48}). This is similar to the discussion presented in Sections 2.2-2.4, except that in the present discussion we will consider a combination of all of the acceleration and loss processes acting simultaneously. The mean rate of change of the momentum is given by (see Equation \ref{eq48})
\begin{equation}
\Big<\frac{dp}{dt}\Big> = D_0 \, m_e c \, \Big<\frac{d\gamma}{dy}\Big>
= D_0 \, m_e c \, (3 + \Atilde - \Btilde \, \gamma^2) \ ,
\label{eq81}
\end{equation}
where we have used the fact that $\gamma = x \equiv p/(m_e c)$ for the ultrarelativistic electrons considered here. We can compute the theoretical maximum Lorentz factor, corresponding to a balance between acceleration and synchrotron losses, by setting $<\!dp/dt\!>=0$, which yields
\begin{equation}
\gamma_{\rm max}^2 =
\frac{\Atilde + 3}{\Btilde}
= \frac{A_{\rm sh} + A_{\rm elec} + 3 D_0}{B_0}
\label{eq82}
\ ,
\end{equation}
where the final result follows from Equations~(\ref{eqA0}), (\ref{eqA0shock}), (\ref{eqA0elec}), and (\ref{eq44}).
Equation~(\ref{eq82}) can be rewritten as
\begin{equation}
\gamma_{\rm max}^2 = \gamma_{\rm elec}^2 + \gamma_{\rm sh}^2
+ \gamma_{\rm stoch}^2
\ ,
\label{eq83}
\end{equation}
where $\gamma_{\rm elec}$ represents the equilibrium Lorentz factor for electrostatic acceleration, given by (see Equation~(\ref{eqElecEquil3}))
\begin{equation}
\gamma_{\rm elec}^2 = \frac{A_{\rm elec}}{B_0}
= \frac{6 \pi q E}{\sigma_{\rm T} B^2}
\ ,
\label{eq84}
\end{equation}
the quantity $\gamma_{\rm sh}$ denotes the equilibrium Lorentz factor for shock acceleration (see Equation~(\ref{eqShockEquil3})),
\begin{equation}
\gamma_{\rm stoch}^2 = \frac{3 D_0}{B_0}
= \frac{18 \pi D_0 m_e c}{\sigma_{\rm T} B^2}
\ .
\end{equation}
and the quantity $\gamma_{\rm stoch}$ represents the equilibrium Lorentz factor for stochastic acceleration, given by (see Equation~(\ref{eqStochEquil3}))
\begin{equation}
\gamma_{\rm sh}^2 = \frac{A_{\rm sh}}{B_0}
= \frac{6 \pi \xi q}{\sigma_{\rm T} B}
\ .
\label{eq85}
\end{equation}
Note that when the shock efficiency parameter $\xi=1$, we find that $\gamma_{\rm sh}$ equals the value of $\gamma_{_{\rm MHD}}$ given by Equation~(\ref{eq8newnew}), as expected, since in this case the shock acceleration occurs at the maximum rate allowed by the synchrotron burnoff condition.

Cerutti et al. (2013) also derived an expression for the maximum Lorentz factor for pure electrostatic acceleration, and it is interesting to compare our result with theirs. They give the maximum Lorentz factor as a function of the particle pitch angle $\theta$ in their Equation~(1), which can be written as
\begin{equation}
\gamma_{\rm elec}^2(\theta) = \frac{4 \pi q E}{\sigma_{\rm T}
B^2 \sin^2\theta}
\ .
\label{eq86}
\end{equation}
In the present paper, we are assuming an isotropic distribution of electron velocities. We can average the Cerutti et al. (2013) result with respect to pitch angle by noting that $<\sin^2\theta> = 2/3$, in which case we obtain our result given by Equation~(\ref{eq84}). Hence the two results for the maximum Lorentz factor for pure electrostatic acceleration are consistent.

Setting $\gamma=\gamma_{\rm max}$ in Equation~(\ref{eq1newnew}) and making use of Equations~(\ref{eq83}) -- (\ref{eq85}), we can obtain an expression for the peak photon energy during the $\gamma$-ray flare, taking into account synchrotron losses as well as stochastic, shock, and electrostatic acceleration. The result obtained is
\begin{equation}
\epsilon_{\rm max} \equiv
\epsilon_{\rm pk}(\gamma_{\rm max})
= \frac{6 \pi \, q \, m_e c^2}{B_{\rm crit} \sigma_{\rm T}}
\left(\xi + \frac{E}{B}
+ \frac{D_0}{D_0^{\rm max}}
\right) \ ,
\label{eq87}
\end{equation}
where we have also made use of Equation~(\ref{eq80}). Equation~(\ref{eq87}) can also be rewritten as
\begin{equation}
\epsilon_{\rm max} = 158 \ {\rm MeV} \
\left(\xi + \frac{E}{B}
+ \frac{D_0}{D_0^{\rm max}}
\right) \ .
\label{eq88}
\end{equation}
We remind the reader that the synchrotron burnoff limit requires that $\xi + D_0/D_0^{\rm max} \le 1$ (see Equation~(\ref{eq118})). Hence, in the absence of electrostatic acceleration, the peak energy cannot exceed 158\,MeV, even in the limiting case $\xi + D_0/D_0^{\rm max} = 1$. This value is too low to explain the highest energy photons emitted during the $\gamma$-ray flares from the Crab nebula, and therefore, in agreement with Cerutti et al. (2012), we find that additional acceleration must be provided by the electric field in order to raise the peak photon energy up to the observed range. The values for $E/B$ obtained using our model are listed in Table~1. In particular, we note that the 2011 April and 2013 March flares require values of $E/B$ close to 2 and 3, respectively.

\subsection{Electrostatic Acceleration Versus Shock Acceleration}

It is interesting to compare the relative contributions from shock acceleration and electrostatic acceleration in the vicinity of the termination shock. We have (see Equations~(\ref{eqpdot}) and (\ref{eq21newnew}))
\begin{equation}
\frac{<\dot p>_{\rm elec}}{<\dot p>_{\rm sh}}
= \frac{E}{\xi B} \ .
\label{eq90}
\end{equation}
The shock acceleration efficiency parameter $\xi=0.1$ in our simulations, so that Equation~(\ref{eq90}) becomes
\begin{equation}
\frac{<\dot p>_{\rm elec}}{<\dot p>_{\rm sh}} = 10 \ \frac{E}{B}
\ ,
\label{eq91}
\end{equation}
In our numerical results for the Crab nebula flares, we generally find that $E/B \sim 0.2-3$, and therefore we can conclude from Equation~(\ref{eq91}) that the first-order (systematic) momentum gain at the termination shock is dominated by electrostatic acceleration, rather than shock acceleration. However, it should be emphasized that despite the negligible role of shock acceleration, the shock nonetheless plays a crucial role in regulating the escape of particles from the acceleration region, and therefore the shock is an essential ingredient in the model.

\subsection{Limits on the Electric Field}
\label{efieldsection}

In order to explain the observed gamma-ray flares from the Crab nebula, our model generally requires a relatively strong electric field, with $E/B \gapprox 1$. Such conditions can be achieved in regions characterized by strong magnetic field gradients, and especially in regions experiencing rapid reconnection (Cerutti  et al. 2012). In this scenario, the magnetic field deep in the reconnection layer can be much smaller than the electric field, and indeed, the magnetic field technically vanishes in the center of the current sheet. Conversely, in the surrounding region, the ambient magnetic field exceeds the electric field. A similar reconnection scenario was considered by Nalewajko et al. (2016) using a time-dependent numerical simulation. They concluded that $E/B \lapprox 1$, which would appear to contradict our requirement that $E/B \gapprox 1$. However, a careful comparison of the two models reveals some subtleties that need to be considered in detail.

In the model of Nalewajko et al. (2016), the value of $E/B$ is computed as a volume average, which is dominated by large portions of the volume containing relatively strong magnetic fields. On the other hand, the one-zone model presented here is not equivalent to a volume average, but is best interpreted as an average weighted by the spatial distribution of the accelerated relativistic electrons, and this distribution is concentrated in the comparatively smaller volume surrounding the shock, where the mean field is smaller. Hence, in our model, the electron-weighted average yields values for $E/B$ that can exceed unity.

\section{DISCUSSION AND CONCLUSION}
\label{sec:discuss}

We have developed and applied a new analytical model for the acceleration and transport of relativistic electrons in pulsar wind termination shocks, including the effects of stochastic wave-particle acceleration, electrostatic acceleration, shock acceleration, particle escape, and energy losses due to synchrotron radiation. This type of scenario has been considered by a number of previous authors as a possible explanation for the $\gamma$-ray flares observed from the Crab nebula using the {\it Fermi}-LAT instrument (Buehler \& Blandford 2014), but we believe that the model developed here is the first to successfully reproduce the $\gamma$-ray flare spectra (see Figure~2). Since energy losses are included in our electron transport equation, the synchrotron flare $\gamma$-ray spectra that we compute are self-consistent. We find that electrostatic acceleration is the main source of energy for the relativistic particles that produce the synchrotron emission in our model, in agreement with Cerutti et al. (2012). The parameters listed in Table~1 show that the electric field required to produce the necessary particle acceleration in our model is moderately enhanced relative to the magnetic field, with $E/B \sim 1-3$, which can be accomplished near the center of the current sheet in a reconnection region (see Section~\ref{efieldsection}). Furthermore, we find that the model is able to roughly reproduce the observed $\gamma$-ray spectra without resorting to Doppler boosting, although including that process would alleviate the power requirements listed in Table~2.

The analytical model considered here differs significantly from the numerical simulations developed by previous authors such as Cerutti et al. (2012, 2013, 2014a) in that we do not explicitly treat the spatial transport, and instead, we focus on qualitatively fitting the $\gamma$-ray spectra using a simplified one-zone model in which the electron energy distribution is interpreted as an average over the acceleration/emission region. In this sense, our work is complementary to the earlier studies. However, the spatial geometry of the problem is treated implicitly through the utilization of a realistic dependence of the escape timescale, $t_{\rm esc}$, on the particle momentum, $p$, as expressed by Equation~(\ref{eq37}). At low energies, the electrons are trapped in the flow, and the escape of the particles is regulated by advection, which sweeps the particles away from the shock. At higher energies, the particle mean-free path is large enough to increase the probability of recycling back into the upstream region, so that the electrons can experience further acceleration. At the highest energies in the model, the Larmor radius is comparable to the termination-shock radius, $r_t \sim 10^{17}\,$cm, and the electrons are able to escape into the outer region of the nebula. The transition between the dominance of the two escape mechanisms occurs at the cross-over Lorentz factor, $\gamma_c$ (Equation~(\ref{eq38})), and the highest-energy electrons accelerated in our model reach a peak Lorentz factor of $\sim 10^{10}$, which is consistent with the Hillas condition for the Crab nebula termination shock radius (see Equation~(\ref{eq35})).

We have also reported computations of the synchrotron afterglow emission produced by the accelerated electrons after they escape into the downstream (outer) cooling region in Figure~5. We predict that the afterglow should be observable for a maximum of approximately 3 weeks. However, this estimate neglects the effects of adiabatic cooling and the decreasing magnetic field strength in the outer region that the escaping particles will experience as that advect through the synchrotron nebula. 

\subsection{X-ray Constraints}

The primary goal of this paper is to develop a new theoretical explanation for the $\gamma$-ray flare spectra observed from the Crab nebula by {\it Fermi}-LAT. However, it is also interesting to determine the level of X-ray emission predicted by our model, since the X-ray spectrum is constrained by previous work. An important characteristic of the {\it Fermi}-LAT flares is that there is no corresponding increase in flux in the radio, optical, UV, or X-ray bands (Abdo et al. 2011; Weisskopf et al. 2013). This observational fact has led many authors to reject the notion of significant Doppler boosting as the underlying cause for the enhanced $\gamma$-ray emission that characterize the flares (Buehler \& Blandford 2014).

Weisskopf et al. (2013) fit the April 2011 $\gamma$-ray flare spectrum using standard spectral analysis software, and extrapolated the fit down to the X-ray region. By employing a statistical pixel analysis routine, they estimated the upper limit on the X-ray flux coincident with the April 2011 flare. In Figure 9, we compare the Weisskopf et al. (2013) upper limits with our calculation of the April 2011 flare spectrum, computed assuming $B = 200\,\mu$G. We find that our model spectrum predicts X-ray fluxes below the upper limits obtained by Weisskopf et al. (2013), which helps to establish further confidence in our model.

\subsection{Magnetic Field Variation}
\label{sec:MagVariation}

The characteristics of the high-energy synchrotron emission produced during the observed $\gamma$-ray flares from the Crab nebula depend sensitively on the strength of the local magnetic field. Rigorous multi-wavelength observational campaigns have concluded that the average strength of the ambient magnetic field in the Crab nebula is $\sim 200 \, \mu$G (Aharonian et al. 2004). Therefore, in the study presented here, we have generally set the field strength using $B = 200 \, \mu$G. However, near the termination shock, compression and reconnection of the turbulent magnetic field can cause a reduction in the local magnetic field strength (Cerutti et al. 2013). It is therefore interesting to reconsider some of our conclusions using a lower estimate for the magnetic field strength $B$. In this section, we present a summary study of the resulting $\gamma$-ray flare spectrum for the April 2011 flare for the case with $B = 100\,\mu$G.

In Figure 8, we plot the resulting theoretical $\gamma$-ray peak flare spectrum obtained when $B = 100\,\mu$G, and compare it with our earlier result, obtained using the canonical field strength $B = 200 \, \mu$G. For the calculation with the reduced field strength $B = 100\,\mu$G, the model parameters are given by $\tilde A = 76$, $\tilde B = 1.75 \times 10^{-19}$, $\tilde C = 18$, $\dot N_0 = 2.56 \times 10^{34} \  {\rm s}^{-1}$, and $\gamma_0 = 10^6$. We observe that the reduced magnetic field strength requires a higher value for the acceleration parameter $\tilde A$ in order to roughly match the observed $\gamma$-ray spectrum. Furthermore, the ratio of electric and magnetic fields in this case increases to $E/B \sim 3$ in order to provide sufficient $\gamma$-ray emission at the highest energies. However, the spectrum also exhibits an excess at lower energies, which suggests that the model with $B = 100 \, \mu$G would probably produce an incorrect photon power-law index for any combination of acceleration and loss parameters. These results tend to support the nominal value $B = 200\,\mu$G that we have adopted in most of our calculations, based on previous observational estimates of the ambient field strength in the Crab nebula.

\subsection{Conclusion}

In agreement with earlier models, we find that shock and stochastic acceleration alone cannot provide sufficient power to explain the $\gamma$-ray luminosity observed during the recent series of Crab nebula flares observed using {\it Fermi}-LAT (e.g. Cerutti et al. 2012). In each case, we conclude that the electrons are primarily accelerated by relatively strong electric fields generated in the region of magnetic reconnection in the vicinity of the pulsar wind termination shock, in agreement with earlier work. Since the model developed here is analytical, and it provides a very flexible and convenient tool that can be used to conduct a broad range of parameter studies. It should be straightforward to port the model into any of the standard data analysis packages in order to perform quantitative fits, which we do not pursue in this paper.

We have assumed here that the electron distribution during the flare peak is close to equilibrium, based on the similarity between the synchrotron loss timescale and the typical $\gamma$-ray flare duration (see Section 1.3), and therefore we have utilized a steady-state model. However, this is clearly a crude approximation of what is intrinsically a time-dependent phenomenon. In future work, we intend to pursue a fully time-dependent simulation, and we also plan to analyze the spectrum resulting from the acceleration of an injected power-law distribution of electrons, rather than the exclusively monoenergetic injection scenario considered here. We will also examine the effects of mild Doppler boosting and anisotropy in the emission beam, which will tend to decrease the power requirements listed in Table~2. We believe that this work will help to guide and inform future studies dedicated to understanding the properties of Crab pulsar magnetosphere and wind.

We are grateful to the anonymous referee for several useful comments that led to improvements in the presentation. P. A. B. would also like to acknowledge support from the Fermi Guest Investigator program during part of this work. J. D. F. and C. D. D. were partially supported by the Chief of Naval Research.

\clearpage

\clearpage

\hoffset=-0.88truein
\begin{deluxetable}{clccccccccccc}
\tabletypesize{\scriptsize}
\tablecaption{Model Parameters\label{tbl-1}}
\tablewidth{0pt}
\tablehead{
\colhead{Flare}
& \colhead{$\sigma_{\textrm{mag}}$}
& \colhead{$\Atilde_{\rm sh}$}
& \colhead{$\Atilde_{\rm elec}$}
& \colhead{$\Btilde$}
& \colhead{$\Ctilde$}
& \colhead{$m_-$}
& \colhead{$D_0 (\rm s^{-1})$}
& \colhead{$\dfrac{E}{B}$}
& \colhead{$w$}
& \colhead{$\dot N_0 (\rm s^{-1})$}
& \colhead{$\gamma_{\rm c}$}
& \colhead{$\gamma_0$}
}
\startdata
September 2007
&0.0802
&3.740
&32.26
&$5.50 \times 10^{-19}$
&10.0
&-0.261
&94.00
&0.862
&3.74
&$4.50 \times 10^{33}$
&$1.12 \times 10^{11}$
&$1 \times 10^{6}$
\\
February 2009
&0.0401
&7.480
&17.52
&$1.10 \times 10^{-18}$
&45.0
&-1.575
&47.00
&0.234
&1.66
&$4.50 \times 10^{38}$
&$5.58 \times 10^{10}$
&$1 \times 10^{6}$
\\
September 2010
&0.0980
&3.060
&32.94
&$4.50 \times 10^{-19}$
&53.0
&-1.347
&114.9
&1.075
&0.58
&$6.00 \times 10^{37}$
&$1.36 \times 10^{11}$
&$1 \times 10^{6}$
\\
April 2011
&0.1026
&2.925
&46.80
&$4.30 \times 10^{-19}$
&15.0
&-0.288
&120.2
&1.600
&1.95
&$8.10 \times 10^{33}$
&$9.21 \times 10^{10}$
&$1 \times 10^{6}$
\\
March 2013
&0.6784
&0.440
&13.56
&$6.50 \times 10^{-20}$
&40.0
&-2.198
&795.4
&3.064
&0.11
&$8.00 \times 10^{35}$
&$5.89 \times 10^{11}$
&$5 \times 10^{8}$
\\
\enddata
\end{deluxetable}

\hoffset=-0.88truein
\begin{deluxetable}{cccccccc}
\tabletypesize{\scriptsize}
\tablecaption{Flare Energy Budgets \label{tbl-2}}
\tablewidth{0pt}
\tablehead{
\colhead{Flare}
& \colhead{$P_{\rm inj} (\rm ergs\,s^{-1})$}
& \colhead{$P_{\rm elec} (\rm ergs\,s^{-1})$}
& \colhead{$P_{\rm sh} (\rm ergs\,s^{-1})$}
& \colhead{$P_{\rm stoch} (\rm ergs\,s^{-1})$}
& \colhead{$P_{\rm syn} (\rm ergs\,s^{-1})$}
& \colhead{$P_{\rm esc} (\rm ergs\,s^{-1})$}
}
\startdata
September 2007
&$3.68 \times 10^{33}$
&$5.58 \times 10^{36}$
&$6.47 \times 10^{35}$
&$5.19 \times 10^{35}$
&$3.36 \times 10^{36}$
&$3.39 \times 10^{36}$
\\
February 2009
&$3.68 \times 10^{38}$
&$3.70 \times 10^{38}$
&$1.58 \times 10^{38}$
&$6.33 \times 10^{37}$
&$1.82 \times 10^{36}$
&$9.50 \times 10^{38}$
\\
September 2010
&$4.91 \times 10^{37}$
&$1.10 \times 10^{38}$
&$1.02 \times 10^{37}$
&$1.01 \times 10^{37}$
&$1.33 \times 10^{36}$
&$1.78 \times 10^{38}$
\\
April 2011
&$6.63 \times 10^{33}$
&$1.03 \times 10^{37}$
&$6.42 \times 10^{35}$
&$6.58 \times 10^{35}$
&$5.54 \times 10^{36}$
&$6.03 \times 10^{36}$
\\
March 2013
&$3.27 \times 10^{38}$
&$1.89 \times 10^{38}$
&$6.17 \times 10^{36}$
&$4.18 \times 10^{37}$
&$4.53 \times 10^{36}$
&$5.60 \times 10^{38}$
\\
\enddata
\end{deluxetable}

\clearpage

\hoffset=0truein

\begin{figure}[ht]
\vspace{0.0cm}
\centering
\includegraphics[height=11cm]{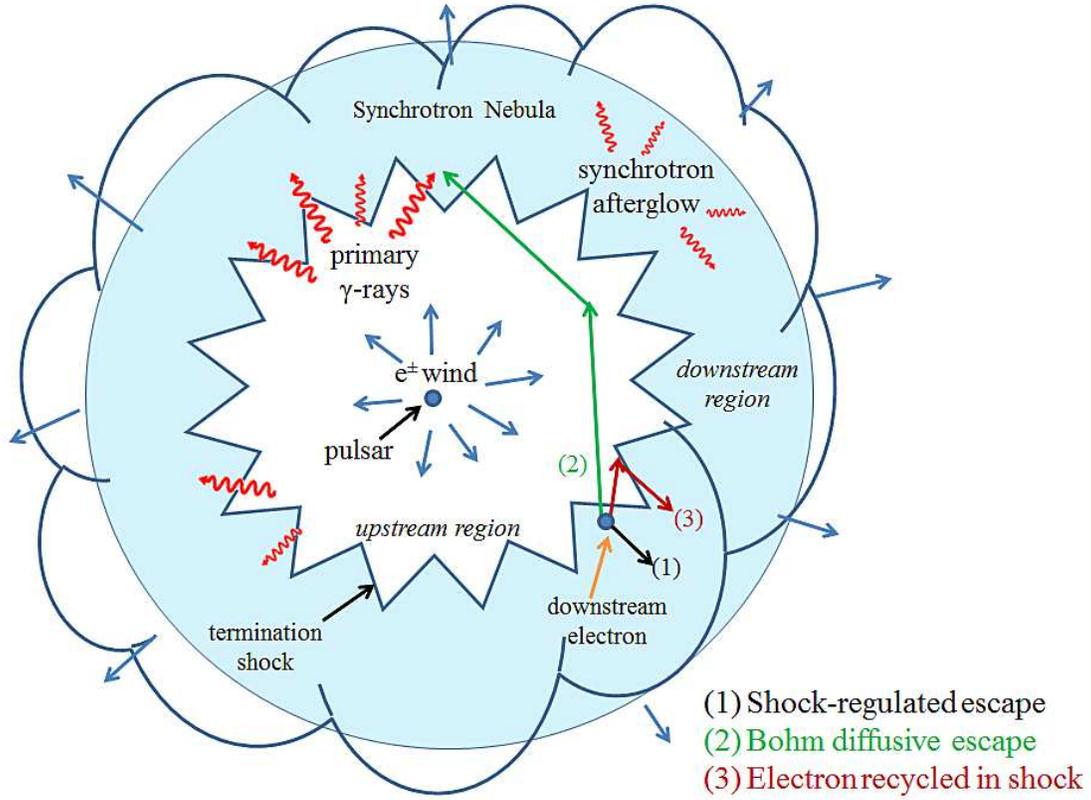}
\caption{Schematic diagram indicating the geometry of the pulsar wind termination shock, and the nature of the particle acceleration and transport processes included in our model.}
\end{figure}

\begin{figure}[ht]
\vspace{0.0cm}
\centering
\includegraphics[height=9cm]{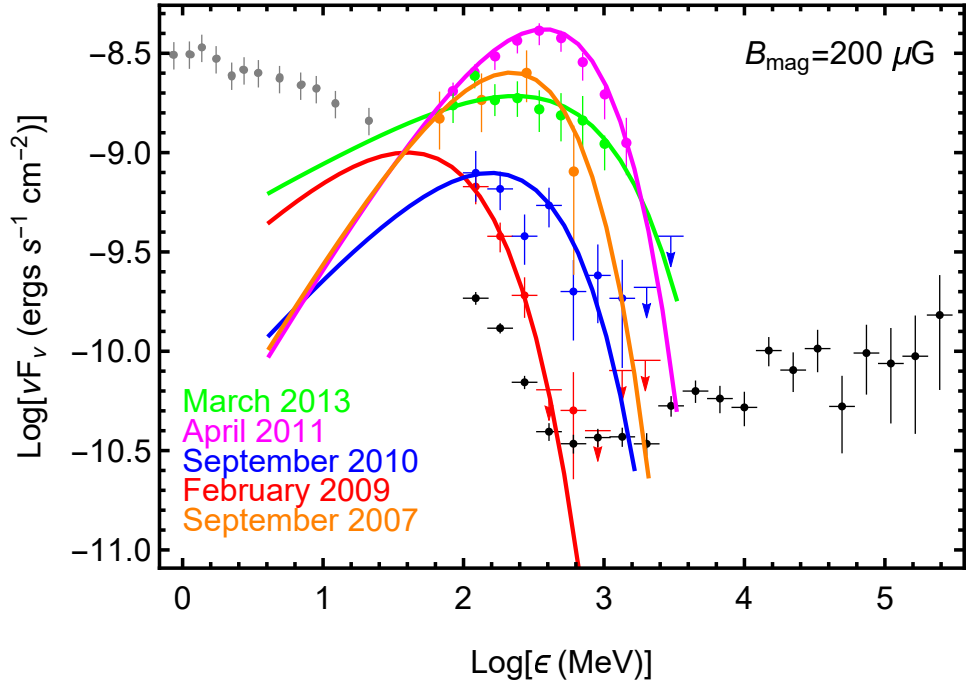}
\caption{Theoretical $\gamma$-ray synchrotron flare spectra, $\nu \mathscr{F}_\nu$, computed using Equation~(\ref{eq72}), plotted as functions of the photon energy $\epsilon$ using the parameters listed in Table~1 (solid lines). The associated electron distributions are computed using Equation~(\ref{eq58}). Also plotted are the corresponding data for each of the $\gamma$-ray flares observed by {\it Fermi}-LAT and {\it AGILE}, taken from Abdo et al. (2011), Buehler et al. (2012), and Buehler \& Blandford (2014).}
\end{figure}

\begin{figure}[ht]
\vspace{0.0cm}
\centering
\includegraphics[height=9cm]{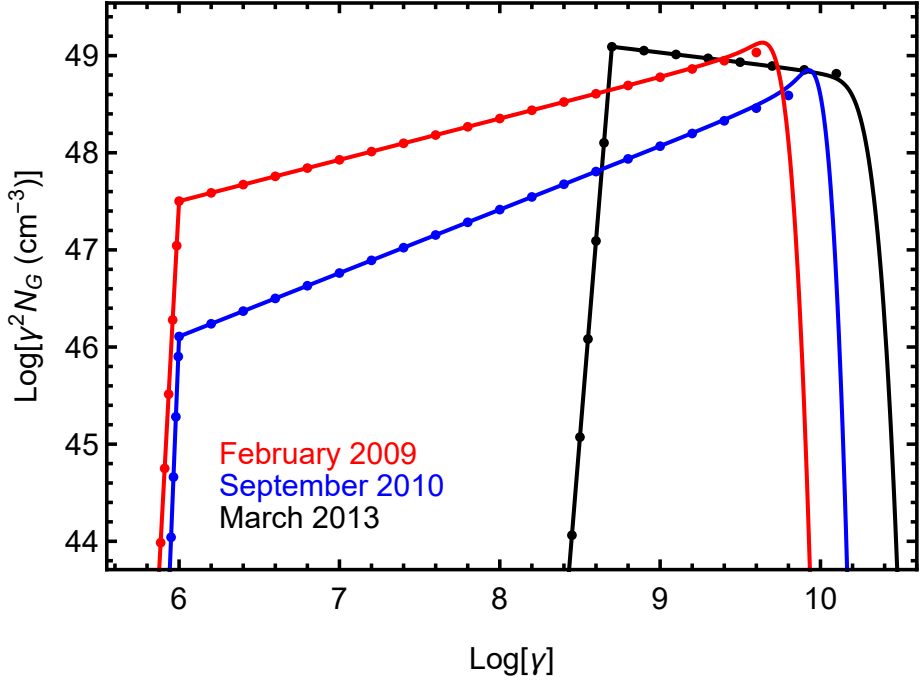}
\caption{The electron number distribution, $\gamma^2 \Ngreen(\gamma,\gamma_0)$, given by the exact solution (Equation~(\ref{eq58}), solid lines) is compared with the approximate broken power-law solution (Equation~(\ref{eq64}), filled circles). Here we consider the electron distributions for the 2009 February, 2010 September, and 2013 March flares, which agree closely with the corresponding power-law distributions up to the exponential turnover created by synchrotron losses. The Lorentz factor of the injected electron, $\gamma_0$, along with the other model parameters for each calculation are indicated in Table~1.}
\end{figure}

\begin{figure}[ht]
\vspace{0.0cm}
\centering
\includegraphics[height=9cm]{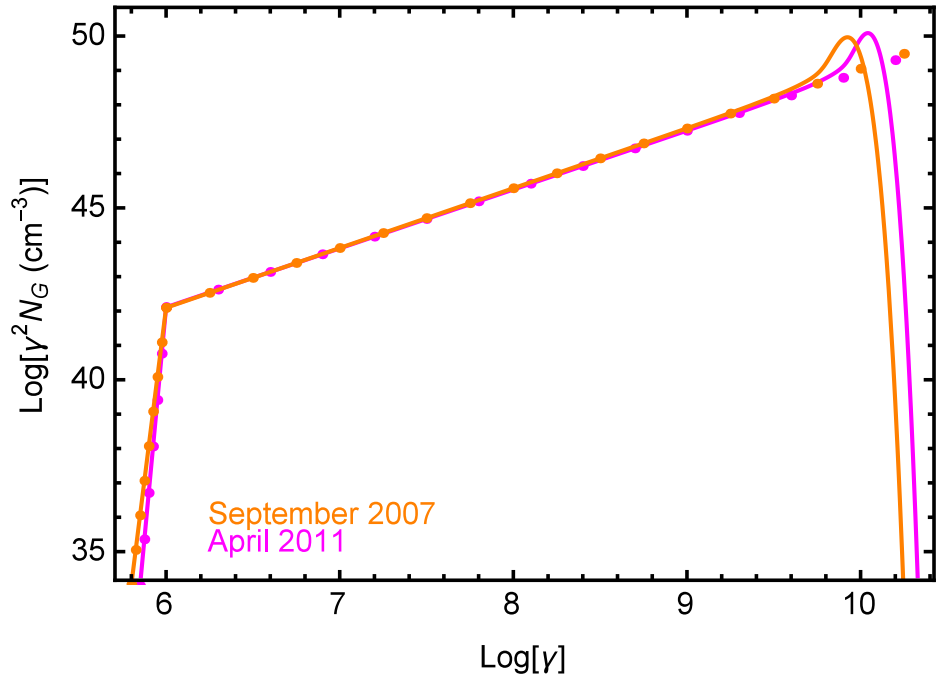}
\caption{Same as Figure~3, except here we plot the electron distributions corresponding to the 2011 April and 2007 September flares. In these two cases, a distinctive particle pile-up occurs at the energy where synchrotron losses produce an exponential turnover. At lower energies, the exact solutions agree with the approximate power-law solution given by Equation~(\ref{eq64}).}
\end{figure}

\begin{figure}
\subfloat{\includegraphics[width = 3in]{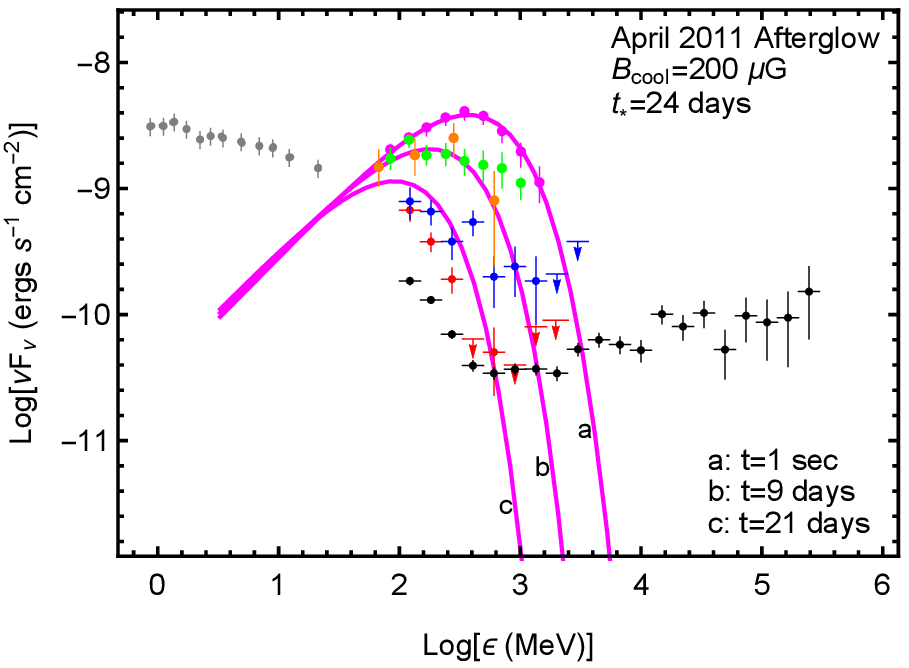}}
\subfloat{\includegraphics[width = 3in]{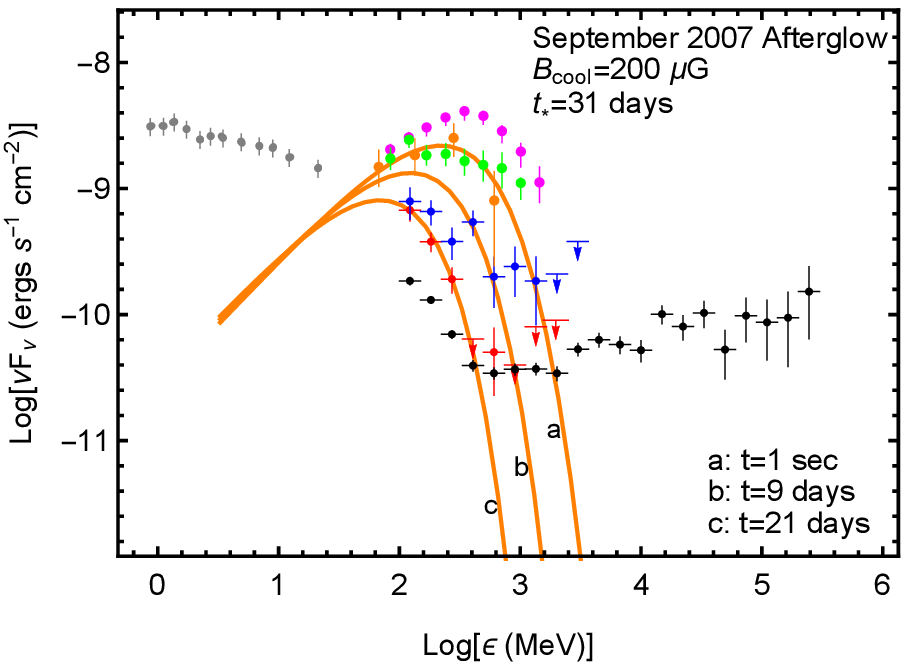}} \ \subfloat{\includegraphics[width = 3in]{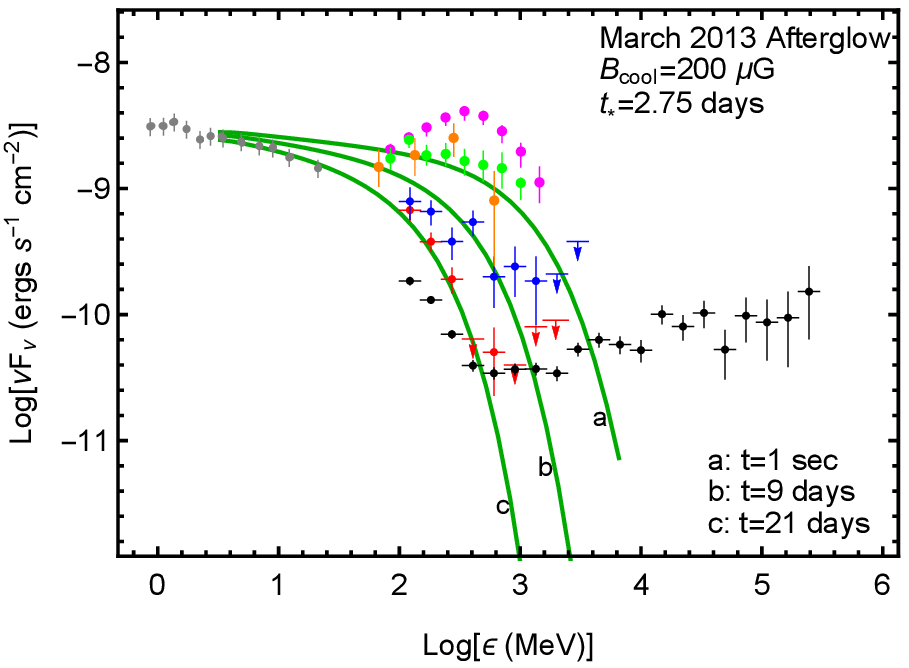}}
\subfloat{\includegraphics[width = 3in]{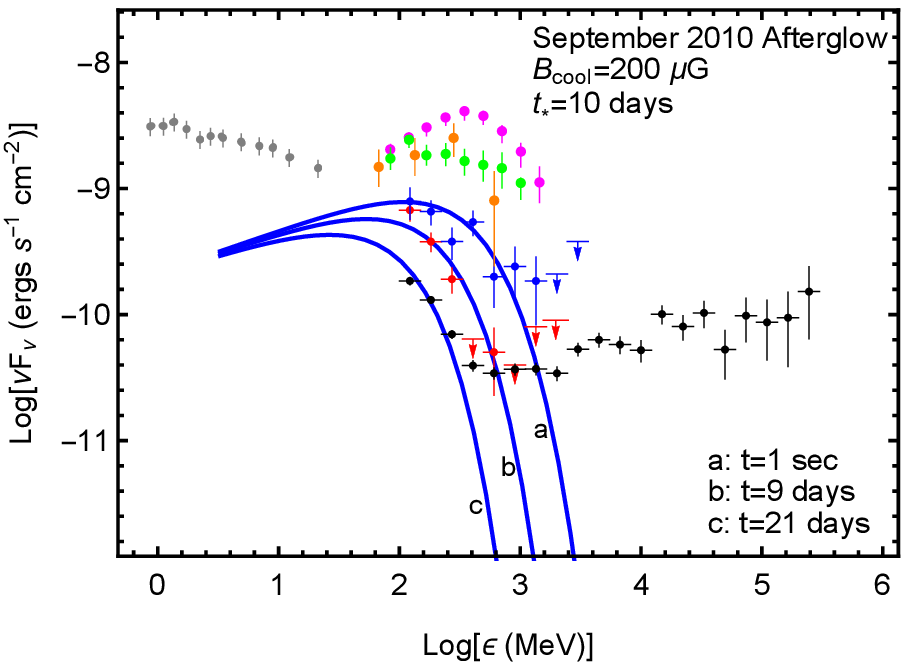}} 
\caption{Theoretical synchrotron afterglow spectra, $\nu \mathscr{F}^{\rm cool}_\nu$, computed using Equation~(\ref{eq109}), plotted as functions of the photon energy $\epsilon$ and the elapsed time $t$ (solid lines) for four of the Crab nebula $\gamma$-ray flares. Also included are the {\it Fermi}-LAT and {\it AGILE} data for all of the flares, as well as the quiescent emission. The magnetic field in the cooling region $B_{\rm cool}=200\,\mu$G as an upper limit. The value of the accumulation timescale $t_*$ is indicated for each flare. The afterglows would be detectable above the quiescent emission for a maximum of about 3 weeks.}
\end{figure}

\begin{figure}[ht]
\vspace{0.0cm}
\centering
\includegraphics[height=9cm]{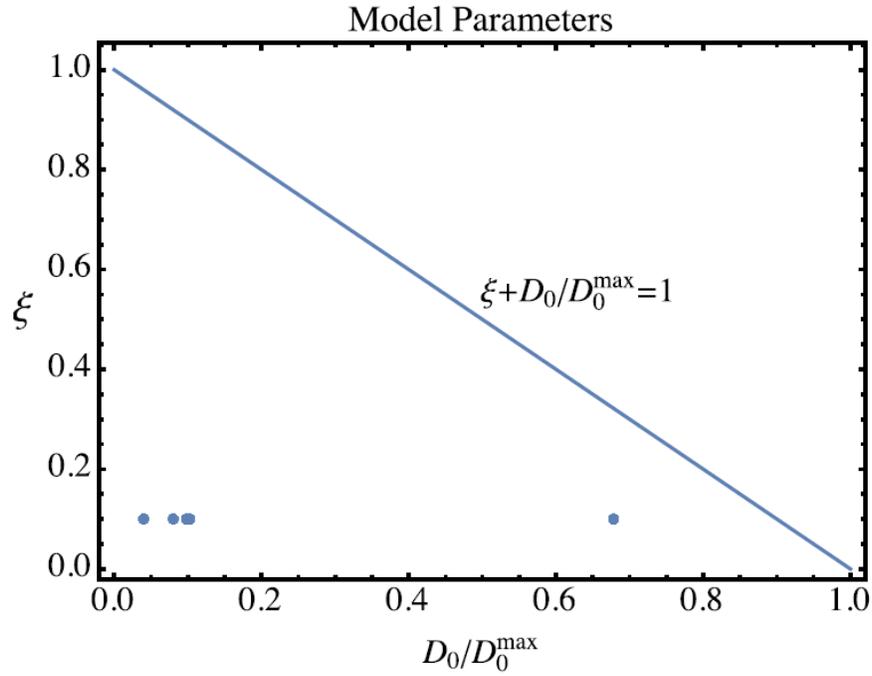}
\caption{Shock acceleration efficiency parameter $\xi$ and momentum diffusion rate constant $D_0/D_0^{\rm max}$ compared with the constraint curve given by Equation~(\ref{eq118}). Each point represents one of the five Crab nebula $\gamma$-ray flares (see Table~1). The synchrotron burnoff limit is satisfied for points below the diagonal line of constraint.}
\end{figure}

\begin{figure}[ht]
\vspace{0.0cm}
\centering
\includegraphics[height=9cm]{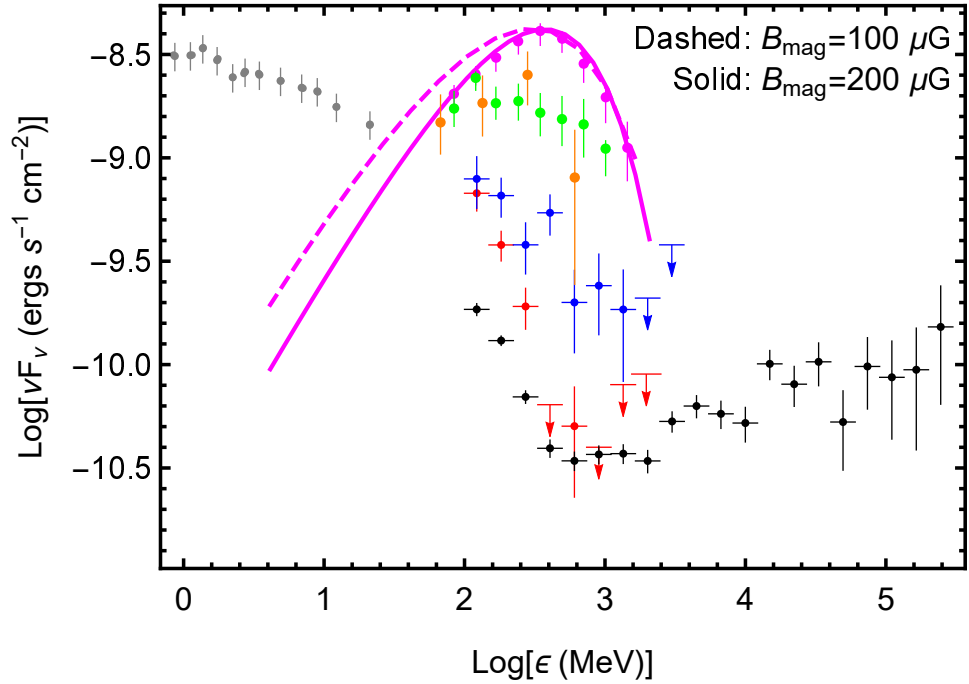}
\caption{Theoretical $\gamma$-ray synchrotron spectrum for the April 2011 flare, computed using an alternative model with reduced magnetic field strength $B=100 \ \mu$G (dashed line), compared with our standard model (solid line). The alternative model parameters are listed in the text in Section~\ref{sec:MagVariation}}
\end{figure}

\begin{figure}[ht]
\vspace{0.0cm}
\centering
\includegraphics[height=9cm]{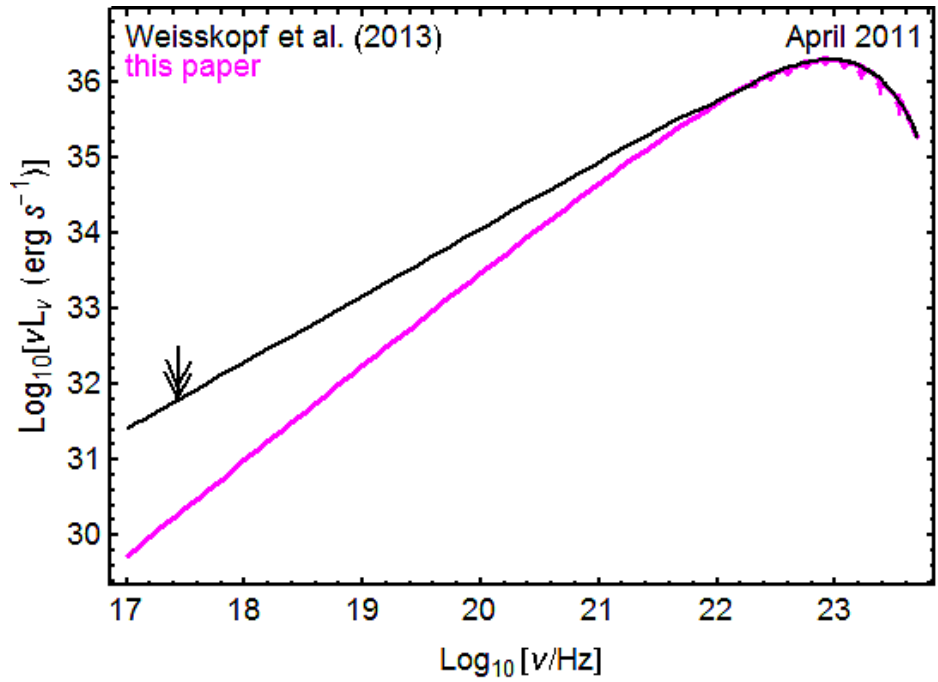}
\caption{Theoretical $\gamma$-ray synchrotron spectrum for the April 2011 flare, computed using the parameters in Table~1. The vertical arrows depict the X-ray upper limits deduced by Weisskopf et al. (2013).}
\end{figure}

{}

\label{lastpage}


\clearpage

\begin{thebibliography}{}

\bibitem{1} Abdo, A., et al. 2011, Science, 331, 739 

\bibitem{2} Abramowitz, M., \& Stegun, I. A. 1970, Handbook of Mathematical Functions, (New York: Dover)

\bibitem{3} Achterberg, A., et al. 2001, MNRAS, 328, 393

\bibitem{4} Aharonian, F., et al. 2004, ApJ, 614, 897 

\bibitem{5} Becker, P. A., \& Begelman, M. C. 1986, ApJ, 310, 534

\bibitem{6} Becker, P. A., Le, T., \& Dermer, C. D. 2006, ApJ, 647, 539

\bibitem{7} Begelman, M.C. 1998, ApJ, 493, 291

\bibitem{8} Buehler, B., \& Blandford, R. 2014, RPPh, 77, 66901

\bibitem{9} Buehler, B., Scargle, J. D., et al. 2012, ApJ, 749, 26

\bibitem{10} Cerutti, B., Uzdensky, D. A., \& Begelman, M.C. 2012, ApJ, 746,148

\bibitem{11} Cerutti, B., Werner, G. R., Uzdensky, D. A., \& Begelman, M. C. 2013, ApJ, 770, 147

\bibitem{12} Cerutti, B., Werner, G. R., Uzdensky, D. A., \& Begelman, M. C. 2014a, ApJ, 782, 104

\bibitem{13} Cerutti, B., Werner, G. R., Uzdensky, D. A., \& Begelman, M. C. 2014b, PhPl, 21, 6501

\bibitem{14} Coroniti, F. V. 1990, ApJ, 349, 538

\bibitem{15} Crusius, A., \& Schlickeiser, R. 1986, A\&A, 164, L16

\bibitem{16} Dermer, C. D., \& Menon, G. 2009, High Energy Radiation from Black Holes: Gamma Rays, Cosmic Rays, and Neutrinos (Princeton, NJ: Princeton Univ. Press)

\bibitem{17} Dermer, C. D., Miller, J. A., \& Li, H. 1996, ApJ, 456, 106

\bibitem{18} Dosch, A., Shalchi, A., \& Tautz, R. C. 2011, MNRAS, 413, 2950

\bibitem{19} Dr\"oge, W., Lerche, I., \& Schlickeiser, R. 1987, A\&A, 178, 252

\bibitem{20} Dr\"oge, W., \& Schlickeiser, R. 1986, ApJ, 305, 909

\bibitem{21} Ellison, D. C., Jones, F. C., \& Reynolds, S. P. 1990, ApJ, 360, 702

\bibitem{22} Gaensler, B. M., \& Slane, P. O. 2006, ARA\&A, 44, 17

\bibitem{23} Gallant, Y. A., \& Achterberg, A. 1999, MNRAS, 305, L6

\bibitem{24} Gallant, Y. A., et al. 1992, ApJ, 391, 73

\bibitem{25} Giacalone, J., \& Jokipii, J. R. 1999, ApJ, 520, 204

\bibitem{26} Hester, J. J. 2008, ARA\&A, 46, 127

\bibitem{27} Heyvaerts, J., Lehner, T., \& Mottez, F. 2012, A\&A, 542, 128

\bibitem{28} Hillas, A. M. 1984, ARA\&A, 22, 425

\bibitem{29} Hussein, M., \& Shalchi, A. 2014, ApJ, 785, 31

\bibitem{30} Jokipii, J. R. 1987, ApJ, 313, 842

\bibitem{31} Kargaltsev O., Cerutti B., Lyubarsky Y., Striani E., 2015, Space Science Reviews, 191, 391

\bibitem{32} Kennel, C. F., \& Coroniti, F. V. 1984, ApJ, 283, 710

\bibitem{33} Komissarov, S. S. 2013, MNRAS, 428, 2459

\bibitem{34} Komissarov, S.S., \& Lyubarsky, Y.E. 2004, MNRAS, 349, 77

\bibitem{35} Komissarov, S. S., \& Lyutikov, M. 2011, MNRAS, 414, 2017

\bibitem{36} Krall, N. A., \& Trivelpiece, A. W. 1986, Principles of Plasma Physics (San Francisco: McGraw-Hill)

\bibitem{37} Lagage, P. O., \& Cesarsky, C. J. 1983, A\&A, 118, 223

\bibitem{38} Lemoine, M., \& Waxman, E. 2009, JCAP, 2009, 9

\bibitem{39} Lyubarsky, Y. E. 2003, MNRAS, 345, 153

\bibitem{40} Lyutikov, M., Balsara, D., \& Matthews, C. 2012, MNRAS, 422, 3118

\bibitem{41} Mayer, M., et al. 2013, ApJ Letters, 775, L37

\bibitem{42} Meyer, M., Horns, D., \& Zechlin, H. S. 2010, A\&A, 523, A2

\bibitem{43} Montani, G., \& Bernardini, M. G. 2014, Physics Letters B, 739, 433

\bibitem{44} Mori, K., Burrows, D.N., Hester, J.J., et al. 2004 ApJ, 609, 186

\bibitem{45} Nalewajko, K., Zrake, J., Yuan, Y., East, W.E., \& Blandford, R.D. 2016, ApJ, 826, 115

\bibitem{46} Olmi, B., Del Zanna, L., Amato, E., \& Bucciantini, N. 2015, MNRAS, 449, 3149

\bibitem{47} Park, B. T., \& Petrosian, V. 1995, ApJ, 446, 699

\bibitem{48} Rees, M. J., \& Gunn, J. E. 1974, MNRAS, 167, 1

\bibitem{49} Reif, F. 1965, Fundamentals of Statistical and Thermal Physics (New York: McGraw-Hill)

\bibitem{50} Rybicki, G. B., \& Lightman, A. P. 1979, Radiative Processes in Astrophysics (New York: Wiley)

\bibitem{51} Schlickeiser, R. 1985 A\&A, 143, 431

\bibitem{52} Sironi, L., Keshet, U., \& Lemoine, M. 2015, SSRv, 191, 519

\bibitem{53} Sironi, L., \& Spitkovsky, A. 2014, ApJ, 783, L21

\bibitem{54} Steinacker, J., \& Schlickeiser, R. 1989, A\&A, 224, 259

\bibitem{55} Striani, E., Tavani, M., Vittorini, V. et al. 2013, ApJ, 765, 52

\bibitem{56} Tavani M., Bulgarelli, A., Vittorini, V., et al. 2011, Sci, 331, 736

\bibitem{57} Uzdensky, D.A., Cerutti, B., \& Begelman, M.C. 2011, ApJ, 737, L40

\bibitem{58} Volpi, D., Del Zanna, L., Amato, E., \& Bucciantini, N. 2008, A\&A, 485, 337

\bibitem{59} Weisskopf, M. C., et al. 2013, ApJ, 765, 56

\bibitem{60} Zank, G. P., Li, G., Florinski, V., et al. 2004, JGR, 109, 4107

\bibitem{61} Zrake, J. 2016, ApJ, 823, 7

\end{thebibliography}
\end{document}